\documentclass[10pt,a4paper,tightenlines]{revtex4}
\usepackage{amssymb,amsmath}
\usepackage{wrapfig}
\usepackage{psfrag}
\usepackage[dvips]{graphicx}
\usepackage{epic,curves}
\newcommand{\ctg}{\mathop{\rm ctg}\nolimits}

\newsavebox{\mirrorEast}
\newsavebox{\mirrorNorth}
\newsavebox{\mirrorWest}
\newsavebox{\mirrorSouth}
\newsavebox{\beamsplSWNO}
\newsavebox{\beamsplSENW}
\newsavebox{\fabryH}
\newsavebox{\fabryV}
\newsavebox{\FabryH}
\newsavebox{\FabryV}
\newsavebox{\abFirst}
\newsavebox{\abSecond}
\newsavebox{\abThird}
\newsavebox{\abFourth}
\newsavebox{\beamsplMain}
\newsavebox{\FabryMich}
\newsavebox{\Speedmeter}

\savebox{\mirrorEast}(2,10){
  \begin{picture}(2,10)
    \thicklines\drawline(2,0)(0,0)(0,10)(2,10)\curve(2,0,1,5,2,10)
  \end{picture}
}

\savebox{\mirrorNorth}(10,2){
  \begin{picture}(10,2)
    \thicklines\drawline(0,2)(0,0)(10,0)(10,2)\curve(0,2,5,1,10,2)
  \end{picture}
}

\savebox{\mirrorWest}(2,10){
  \begin{picture}(2,10)
    \thicklines\drawline(0,0)(2,0)(2,10)(0,10)\curve(0,0,1,5,0,10)
  \end{picture}
}

\savebox{\mirrorSouth}(10,2){
  \begin{picture}(10,2)
    \thicklines\drawline(0,0)(0,2)(10,2)(10,0)\curve(0,0,5,1,10,0)
  \end{picture}
}

\savebox{\beamsplSWNO}(10,10){
  \begin{picture}(10,10)
    \thicklines\drawline(0,0)(10,0)(10,10)(0,10)(0,0)(10,10)
  \end{picture}
}

\savebox{\beamsplSENW}(10,10){
  \begin{picture}(10,10)
    \thicklines\drawline(10,0)(10,10)(0,10)(0,0)(10,0)(0,10)
  \end{picture}
}

\savebox{\fabryH}(20,10){
  \begin{picture}(20,10)
    \put(10,0){\usebox{\mirrorEast}}
    \put(18,0){\usebox{\mirrorWest}}
    \thicklines
    \put(10,5){\vector(-1,0){10}}\put(0,4){\makebox(0,0)[lt]{$x_3$}}
  \end{picture}
}

\savebox{\fabryV}(10,20){
  \begin{picture}(10,20)
    \put(0,10){\usebox{\mirrorNorth}}
    \put(0,18){\usebox{\mirrorSouth}}
    \thicklines
    \put(5,10){\vector(0,-1){10}}\put(6,0){\makebox(0,0)[lb]{$x_4$}}
  \end{picture}
}

\savebox{\FabryH}(30,10){
  \begin{picture}(30,10)
    \put(0,0){\usebox{\mirrorEast}}
    \put(18,0){\usebox{\mirrorWest}}
    \thicklines
    \put(20,5){\vector(1,0){10}}\put(30,4){\makebox(0,0)[rt]{$x_e$}}
  \end{picture}
}

\savebox{\FabryV}(10,30){
  \begin{picture}(10,30)
    \put(0,0){\usebox{\mirrorNorth}}
    \put(0,18){\usebox{\mirrorSouth}}
    \thicklines
    \put(5,20){\vector(0,1){10}}\put(6,30){\makebox(0,0)[lt]{$x_n$}}
  \end{picture}
}

\savebox{\abFirst}(10,10){
  \begin{picture}(10,10)
    \thicklines
    \put(2.5,6){\vector(1,0){5}}\put(5,7){\makebox(0,0)[cb]{$a_n$}}
    \put(7.5,4){\vector(-1,0){5}}\put(5,3){\makebox(0,0)[ct]{$c_e$}}
  \end{picture}
}

\savebox{\abSecond}(10,10){
  \begin{picture}(10,10)
    \put(4,2.5){\vector(0,1){5}}\put(3,5){\makebox(0,0)[rc]{$a_e$}}
    \put(6,7.5){\vector(0,-1){5}}\put(7,5){\makebox(0,0)[lc]{$c_n$}}
  \end{picture}
}

\savebox{\abThird}(10,10){
  \begin{picture}(10,10)
    \put(7.5,6){\vector(-1,0){5}}\put(5,7){\makebox(0,0)[cb]{$c_w$}}
    \put(2.5,4){\vector(1,0){5}}\put(5,3){\makebox(0,0)[ct]{$a_w$}}
  \end{picture}
}
\savebox{\abFourth}(10,10){
  \begin{picture}(10,10)
    \put(4,7.5){\vector(0,-1){5}}\put(3,4){\makebox(0,0)[rc]{$c_s$}}
    \put(6,2.5){\vector(0,1){5}}\put(7,4){\makebox(0,0)[lc]{$a_s$}}
  \end{picture}
}

\savebox{\beamsplMain}(25,20){
  \begin{picture}(25,20)
    \put(0,0){\usebox{\beamsplSWNO}}
    \put(20,0){\usebox{\abFirst}}
    \put(0,10){\usebox{\abSecond}}
  \end{picture}
}

\savebox{\FabryMich}(60,50){
  \begin{picture}(60,50)
    \put(0,0){\usebox{\beamsplMain}}
    \put(30,0){\usebox{\FabryH}}
    \put(0,20){\usebox{\FabryV}}
  \end{picture}
}

\savebox{\Speedmeter}(120,120){
  \begin{picture}(120,120)
    \thicklines
    \put(0,10){\usebox{\abThird}}
    \put(10,0){\usebox{\abFourth}}
    \put(10,10){\usebox{\beamsplMain}}
    \drawline(10,30)(20,40)\drawline(40,10)(50,20)
    \put(40,30){\usebox{\beamsplSENW}}
    \put(39,29){\makebox(0,0)[rt]{\sf PBS}}
    \put(55,36){\vector(1,0){5}}\put(57.5,39){\makebox(0,0)[cc]{$a_e^{(0^o)}$}}
    \put(55,43){\vector(1,0){5}}\put(57.5,46){\makebox(0,0)[cc]{$b_n^{(90^o)}$}}
    \put(60,34){\vector(-1,0){5}}\put(57.5,31){\makebox(0,0)[cc]{$c_n^{(0^o)}$}}
    \put(60,28){\vector(-1,0){5}}\put(57.5,25){\makebox(0,0)[cc]{$b_e^{(90^o)}$}}
    \drawline(67,27)(72,27)(72,43)(67,43)(67,27)(72,43)
    \put(70,25){\makebox(0,0)[ct]{$\lambda/4$}}
    \put(80,36){\vector(1,0){5}}\put(82.5,39){\makebox(0,0)[cc]{$a_e^{\circlearrowright}$}}
    \put(80,43){\vector(1,0){5}}\put(82.5,46){\makebox(0,0)[cc]{$b_n^{\circlearrowleft}$}}
    \put(85,34){\vector(-1,0){5}}\put(82.5,31){\makebox(0,0)[cc]{$c_n^{\circlearrowleft}$}}
    \put(85,28){\vector(-1,0){5}}\put(82.5,25){\makebox(0,0)[cc]{$b_e^{\circlearrowright}$}}
    \put(90,30){\usebox{\FabryH}}
    \put(44,47.5){\vector(0,1){5}}\put(43,50){\makebox(0,0)[rc]{$a_n^{{(0^o)}}$}}
    \put(44,55){\vector(0,1){5}}\put(43,57.5){\makebox(0,0)[rc]{$b_e^{{(90^o)}}$}}
    \put(46,52.5){\vector(0,-1){5}}\put(47,50){\makebox(0,0)[lc]{$c_e^{(0^o)}$}}
    \put(46,60){\vector(0,-1){5}}\put(47,57.5){\makebox(0,0)[lc]{$b_n^{(90^o)}$}}
    \drawline(37,65)(52,65)(52,70)(37,70)(37,65)(52,65)(37,70)
    \put(54,67.5){\makebox(0,0)[lc]{$\lambda/4$}}
    \put(44,75){\vector(0,1){5}}\put(43,77.5){\makebox(0,0)[rc]{$a_n^{\circlearrowright}$}}
    \put(44,82.5){\vector(0,1){5}}\put(43,85){\makebox(0,0)[rc]{$b_e^{\circlearrowleft}$}}
    \put(46,80){\vector(0,-1){5}}\put(47,77.5){\makebox(0,0)[lc]{$c_e^{\circlearrowleft}$}}
    \put(46,87.5){\vector(0,-1){5}}\put(47,85){\makebox(0,0)[lc]{$b_n^{\circlearrowright}$}}
    \put(40,90){\usebox{\FabryV}}
  \end{picture}
}

\begin{document}

\title{Sensitivity limitations in optical speed meter topology of gravitational-wave antennae.}

\author{S.L.Danilishin}
\affiliation{Department of Physics, Moscow State University, Moscow 119992, Russia,\\
e-mail: stefan@hbar.phys.msu.ru}

\begin{abstract}
The possible design of QND gravitational-wave detector based on speed meter principle is considered with respect to optical losses. The detailed analysis of speed meter interferometer is performed and the ultimate sensitivity that can be achieved is calculated. It is shown that unlike the position meter signal-recycling can hardly be implemented in speed meter topology to replace the arm cavities as it is done in signal-recycled detectors, such as GEO 600. It is also shown that speed meter can beat the Standard Quantum Limit (SQL) by the factor of $\sim 3$ in relatively wide frequency band, and by the factor of $\sim 10$ in narrow band. For wide band detection speed meter requires quite reasonable amount of circulating power $\sim 1$ MW. The advantage of the considered scheme is that it can be implemented with minimal changes in the current optical layout of LIGO interferometer.
\end{abstract}
\maketitle

\section{Introduction}

It is a well known fact that in order to detect gravitational wave
very high precision devices are needed. The current technological
progress in this area of knowledge suggests that detector will
achieve the level of sensitivity where its quantum behavior will
play the main role.

Sensitivity of the third-generation interferometric gravitational
wave detectors is planned to be higher than Standard Quantum Limit
(SQL) \cite{67a1eBr,76a1eKhVo}. Therefore, to overcome the SQL
one needs to monitor not the coordinate of the detector probe
body, as in contemporary detectors, but the observable that is not
perturbed by the measurement. Such an observable was called
Quantum Non-Demolition (QND) observable
\cite{77a1eBrKhVo,78ThDrCaZiSa,89BookeVo,92BookBrKh}. One needs to
monitor this kind of observable because, if there were no force
acting upon the probe object the observable value will remain
unperturbed after the measurement. It means that there is no back
action and, therefore, no SQL, limiting the sensitivity. In order
to detect external action on the object it is reasonable to
measure its integral of motion that is a QND-observable at the
same time.
For the free mass, its momentum can be that QND
observable, but the realization of momentum measurement is not an
easy task. In the article \cite{90a1BrKh} it was proposed to
measure velocity of a free mass instead of momentum. Though it is
not a QND observable and is perturbed during the measurement, it
returns to the initial value after the measurement, and therefore
can be used to beat the SQL. The device that measures object
velocity is agreed to be called speed meter.

The first realization of gravitational-wave detector based on
speed meter principle was proposed in article \cite{90a1BrKh}. In
this article authors suggested to measure the velocity of the free
mass placed in the gravitational wave field. The analysis of the
scheme has shown that measuring of velocity instead of body
coordinate allows to cancel the back action noise and, therefore,
significantly increase scheme sensitivity. The measuring system
was presented by two coupled microwave cavities which coupling
constant had to be chosen so that phase shift proportional to
coordinate was fully compensated, while the output signal
contained information about the probe mass velocity only.

Later, in article \cite{96a1KhLe} the possibility to apply speed
measurement technique to interferometric gravitational-wave
detectors has been analyzed. It was suggested to attach to the
probe bodies of the detector small rigid Fabry-Perot cavities.
These cavities, fully transparent for light at certain frequency
$\omega_\circ$ when immobile, introduce phase shift to the output
light due to Doppler effect. Measuring this phase shift it is
possible to measure probe body velocity. It was shown in this
article that speed meter can potentially beat the SQL if pumping
power is larger than one for SQL limited position meter: $W>W_{\rm
SQL}$.

The realization of speed meter based on two microwave coupled
resonators, suggested in \cite{90a1BrKh} was considered in article
\cite{00a1BrGoKhTh}. This speed meter was proposed to be attached
to the detector probe body to measure its velocity. It was
demonstrated that it is feasible, with current technology, to
construct such a speed meter that beats the SQL in a wide
frequency band by a factor of $2$. It was also proposed a possible
design of speed meter for optical frequency band. This design
demands construction of four large scale cavities instead of two
as in traditional LIGO detector. The disadvantage of this scheme,
common for all speed meters, is the extremely high power
circulating in cavities.

The further comprehensive analysis of the scheme proposed in
\cite{00a1BrGoKhTh} was carried in \cite{02a1Pu}. It was shown
that {\it in principle} the interferometric speed meter can beat
the gravitational-wave standard quantum limit (SQL) by an
arbitrarily large amount, over an arbitrarily wide range of
frequencies, and can do so without the use of squeezed vacuum or
any auxiliary filter cavities at the interferometer's input or
output. However, {\it in practice}, to reach or beat the SQL, this
specific speed meter requires exorbitantly high input light power.
The influence of losses on the speed meter sensitivity is also
analyzed and it was shown that optical losses in considered scheme
influence the sensitivity at low frequencies.

In articles \cite{02a1PuCh,03a1Ch} more practical schemes of large
scale interferometric speed meters based on Sagnac effect \cite{Sagnac1913}
were analyzed and proposed for use as third generation LIGO
detectors possible design. It should be noted that use of Sagnac interferometers in gravitational-wave detection was suggested earlier in the works \cite{Drever1983,Byer1996,Byer1999,Byer2000}. In  \cite{02a1PuCh,03a1Ch} it was shown that to lower the value of
circulating power one should use the squeezed vacuum input with
squeezing factor of $\sim 0.1$, and variational output detection
\cite{95a1VyZu}. Then it is possible to beat the SQL by the factor
of $\sim \sqrt{10}$. The analyzed schemes are based on using
either three large scale Fabry-Perot cavities \cite{02a1PuCh} or
ring cavities and optical delay lines \cite{03a1Ch}. The
comprehensive analysis of the above mentioned schemes including
optical losses had been carried out.

Another variant of Sagnac based speed meter was proposed in
\cite{02a2Kh}. This design requires little changes in initial LIGO
equipment and seems to be a good candidate for implementation. The
scheme of interferometer proposed in the article mentioned above
is considered in this paper.  We analyze here how this meter will
behave when there are optical losses in interferometer mirrors,
and obtain the optimal value of circulating power and how it
depends on cavity parameters.

This article is organized as follows: In section \ref{SRPMvsSRSM} we compare sensitivities of signal-recycled position meter and speed meter, and demonstrate that the last one has worse sensitivity at the same level of pumping power if signal recycling is applied. In section \ref{Sec2} we
consider the simple scheme of speed meter with single lossy
element in order to study how optical losses influence the scheme
sensitivity.   In section \ref{Sec3} we evaluate sensitivity of more
realistic speed meter scheme proposed in \cite{02a2Kh} and obtain
the optimal values of parameters that minimize influence of noise
sources. In section \ref{Sec4} we summarize our results. Section
\ref{Sec5} is devoted to acknowledgements.

\section{Simple Sagnac speed meter scheme with optical losses}\label{Sec2}
\subsection{Input-output relations for simple speed meter
scheme}

\begin{wrapfigure}[26]{l}{0.48\textwidth}
\begin{center}
\psfrag{Aw}{$a_w$}\psfrag{As}{$a_s$}\psfrag{An}{$a_n$}\psfrag{Ae}{$a_e$}
\psfrag{Bw}{$b_w$}\psfrag{Bs}{$b_s$}\psfrag{Bn}[lc]{$b_e$}\psfrag{Be}{$b_n$}
\psfrag{Gn}{$g_n$}\psfrag{Ge}{$g_e$}\psfrag{Xn}{$x_n$}\psfrag{Xe}{$x_e$}\psfrag{PRM}[lb]{PRM}\psfrag{SRM}{SRM}
\includegraphics[width=0.48\textwidth]{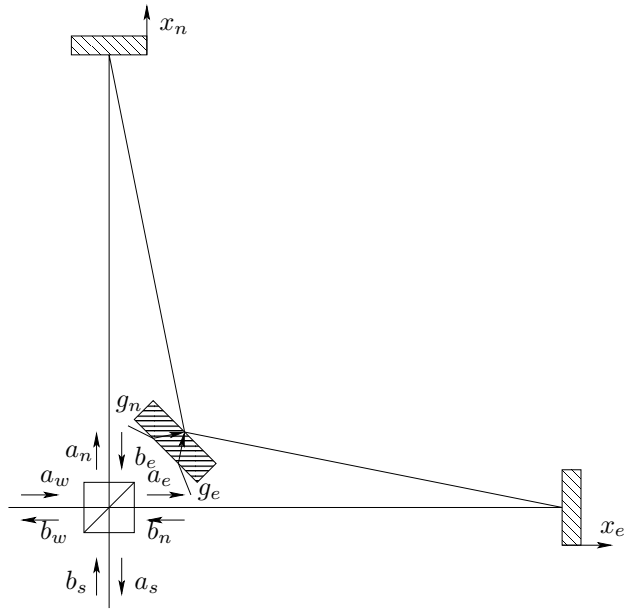}\\
\caption{Simplified scheme of speed
meter.}\label{fig2}
\end{center}
\end{wrapfigure}
In this section we will analyze the action of simple speed meter
scheme based on Sagnac effect to estimate how optical losses affect its sensitivity. This
scheme is the model that is quite easy for analysis and yet contains all features
specific for speed meter. It differs from real, complicated
schemes only by the fact that there is no bending up of noise
curve at high frequencies that arises due to the finite bandwidth
of Fabry-Perot cavities being used in real speed meter
interferometers. Therefore, it seems convenient to investigate
this simple scheme before analyzing complicated speed meters
designed for LIGO.

The scheme we want to consider is presented in Fig. \ref{fig2}.
Its action can be described as follows: the light beam from laser
enters the scheme from the "western" side of the figure and is
divided by the beam-splitter into two beams propagating in
"northern" and "eastern" directions correspondingly. Each beam is
reflected sequentially from two movable end mirrors. Between these
reflections beams are reflected from central mirror.
The end mirrors we suppose to be ideally reflective, while the central
mirror has the amplitude reflectivity $r$ and transmittance $t$.
We suppose that all optical losses are concentrated in the central mirror and therefore can be expressed by a single parameter $t$. It can be shown that for gravitational-wave detectors this approximation is fulfilled pretty well.

After three above mentioned reflections the light
beams return to the beam-splitter where they are mixed up. We
suppose that arm lengths are chosen so that all the light power
returns to the laser. Then photo-detector located in the
"southern" part of the figure will register only vacuum
oscillations whose phase is modulated by the end mirrors motion.

The input light can be presented as a sum of classical pumping
wave with frequency $\omega_\circ$ and sideband quantum
oscillations. The electric field strength of incident wave can be
written as
\begin{equation}\label{EMFquant}
  \hat E(t) = \ell(\omega_0)\left\{Ae^{-i\omega_ot}+A^*e^{i\omega_ot}\right\}+\int\limits^\infty_0\ell(\omega)\left[
    \hat a(\omega)e^{-i\omega t} + \hat a^\dagger(\omega)e^{i\omega t}
  \right]\frac{d\omega}{2\pi}\,,
\end{equation}
where $A$ is classical wave quadrature amplitude, $\hat a(\omega)$
is quantum fluctuations sideband operator, and
$\ell(\omega)=\sqrt{\dfrac{2\pi\hbar\omega}{{\cal A}c}}$ is
normalization factor, and ${\cal A}$ is the beam cross section.
Using this formalism we can write down input-output relations for
the scheme. The incident wave quadrature amplitude and
fluctuations we will denote as $A$ and $\hat a_w$ as they come
from the "western" side of the figure. Zero oscillations entering
the scheme from the "southern" side we will denote as $\hat a_s$.
After the beam-splitter we will have two light beams propagating
in "northern" and "eastern" directions. These beams values we will
mark by indices "n" and "e" correspondingly. Moreover, there is a
transparent central mirror that is the source of additional noises
that are denoted as $\hat g_n$ and $\hat g_e$ (see Fig.
\ref{fig2}).

Let denote the beam that travel in the interferometer arm for the first time as "primary" beam, while the same beam that has left its first arm and entered the second one we will denote as "secondary" beam. Due to absorption the powers of "primary" and "secondary" beams relate to each other as $W_{secondary}/W_{primary}=r^2=1-\alpha_{loss}^2\,,$ where $\alpha_{loss}$ is the interferometer absorption coefficient. This coefficient is quite small and in practice can be neglected for the values of classical powers but, in principle, there is an opportunity to compensate these losses by introducing some additional power into arms. This opportunity can be taken into account by assuming $W_{secondary}/W_{primary}=\eta^2\,,$ where $\eta$ is some coefficient that is equal to $1-\alpha_{loss}^2$ without additional pumping, and can be larger with it. We will assume that $\eta=1$ as optical losses in considered schemes are supposed to be very small ($\alpha_{loss}\sim 10^{-5}$). Then, output wave can be
written as (see Appendix \ref{App1})
\begin{equation}\label{SimpleOutput}
  \hat b_s=\beta_{input}\hat a_s + \beta_{loss}\hat g_s+{\cal
  K}_{simple}x_-\,,
\end{equation}
where $$\beta_{input} = ire^{4i\omega\tau}$$ is the coefficient
that characterizes input fluctuations,
$$\beta_{loss} = -ite^{2i\omega\tau}$$ is the coefficient that
characterizes fluctuations arising due to losses, $${\cal
  K}_{simple} = -2\kappa(\omega)A(ire^{i(\omega_\circ+3\omega)\tau}-\eta e^{i(\omega_\circ+\omega)\tau})\,,$$
is the simple scheme coupling constant, where $\kappa(\omega) =
\sqrt{\omega\omega_\circ}/c$, $\tau$ is the time light needs to
come from the beam-splitter to the end mirror, and
$x_-(\omega_\circ-\omega) = \dfrac{x_e-x_n}{2}$ is the difference
of the end mirrors coordinates in frequency domain. Here $\hat g_s
= \dfrac{i\hat g_e-\hat g_n}{\sqrt2}$ stands for the  loss noises recalculated to
output.

\subsection{Quantum noise spectral density}

In order to evaluate the scheme sensitivity we need to calculate
spectral density of total quantum noise. Suppose mirrors dynamics
can be described by free body equation of motion, \textit{i. e.}
$$m\ddot x=F\,.$$ Here $F=F_{\rm GW}+F_{fluct}\,,$
where $F_{\rm GW}$ is the gravitational-wave force measured,
$F_{fluct}$ is the fluctuational force that arises due to
radiation pressure of the incident light, and $x = x_{\rm
GW}+x_{fluct}$ is the mirror displacement that consists of
displacement due to gravitational wave action and noisy part that
arises due to incident light phase fluctuations. The total noise
spectral density then can be written as
\begin{equation}\label{simplStotal}
  S(\Omega)=S_F(\Omega)+m^2\Omega^4S_x(\Omega)-2m\Omega^2\Re([S_{xF}(\Omega)]\,,
\end{equation}
where $\Omega$ is observation frequency,  and
$S_x(\Omega)$,  $S_F(\Omega)$, $S_{xF}(\Omega)$ are spectral
densities of fluctuational mirror displacement $x_{fluct}$, fluctuational radiation pressure force $F_{fluct}$, and their cross-correlation correspondingly.

In our specific case these spectral densities are equal to (see
Appendix \ref{App1})
\begin{subequations}\label{simpleSpDens}
\renewcommand{\theequation}{\theparentequation.\arabic{equation}}
\begin{equation}
 S_F(\Omega)=\dfrac{8\hbar\omega_\circ W}{c^2}\cdot(1-r\cos(2\Omega\tau))\,,
\end{equation}
\begin{equation}
  S_x(\Omega)=\dfrac{\hbar c^2}{16\omega_\circ
 W\sin^2\Psi}\cdot\dfrac{1}{1+r^2-2r\cos(2\Omega\tau)}\,,
\end{equation}
\begin{equation}
  S_{xF}(\Omega)=-\frac\hbar2\ctg\Psi\,.
\end{equation}
\end{subequations}
Here $W$ is the pumping power at the end
mirrors, and $\Psi $ is the homodyne
angle that allows to minimize the value of
(\ref{simplStotal}) significantly. This angle is chosen so that one measures not
the amplitude or phase quadrature component but their mixture.
This principle provides the basis for variational measurement
technique \cite{95a1VyZu,97a1VyLa}. Here $\Psi$ is one of the
optimization parameters that allows to overcome the SQL being
chosen in the proper way.

Suppose that signal varies slowly compared to the scheme
characteristic time $\tau$, and, therefore
\begin{subequations}\label{NBconditions1}
\renewcommand{\theequation}{\theparentequation.\arabic{equation}}
\begin{equation}
  \Omega\tau\ll1\,,
\end{equation}
\begin{equation}
  \Omega\ll\frac{2\pi}{\tau}\,,
\end{equation}
\end{subequations}
where $\tau=L/c$, and $L$ is the distance between the central and end mirrors. These assumptions we will call
\textit{narrow-band approximation}. Later we will see that for
real detectors this approximation works pretty well.

\begin{wrapfigure}[18]{l}{0.5\textwidth}
\begin{center}
\psfragscanon
\psfrag{W}[lt]{$\Omega\tau$}\psfrag{xi}[cb]{$S/S_{\rm
SQL}$}\psfrag{n1}[lb]{\footnotesize $P_{simple}=1$}\psfrag{n2}[lb]{\footnotesize $P_{simple}=10$}
\psfrag{n3}[lb]{\footnotesize $P_{simple}=100$}
\includegraphics[width=0.4\textwidth]{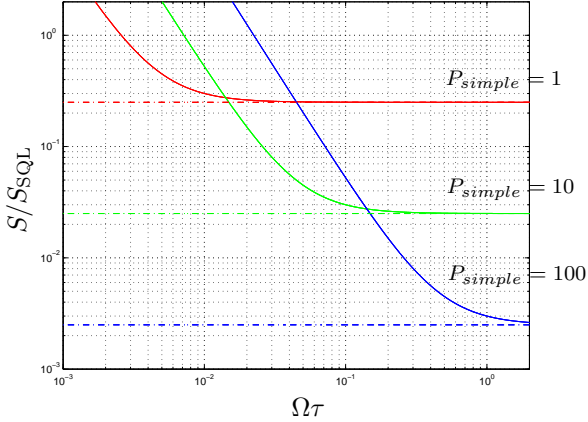}\\
\caption{Typical curves for $\xi^2=S/S_{\rm SQL}$ at different
$P_{simple}$.}\label{fig3}
\end{center}
\end{wrapfigure}

Using this approximation we can rewrite (\ref{simpleSpDens}) as:
\begin{subequations}
\renewcommand{\theequation}{\theparentequation.\arabic{equation}}
\begin{equation}
 S_F(\Omega)=\dfrac{8\hbar\omega_\circ W}{c^2}\cdot(1-r+2r(\Omega\tau)^2)\,,
\end{equation}
\begin{equation}
  S_x(\Omega)=\dfrac{\hbar c^2}{16\omega_\circ
 W\sin^2\Psi}\cdot\dfrac{1}{(1-r)^2+4r(\Omega\tau)^2}\,,
\end{equation}
\begin{equation}
  S_{xF}(\Omega)=-\frac\hbar2\ctg\Psi\,.
\end{equation}
\end{subequations}

If we substitute the obtained expressions to (\ref{simplStotal})
and divide by the value of radiation pressure noise SQL spectral density for the free mass
$S_{\rm SQL}=\hbar m\Omega^2$, we will obtain the factor by which
speed meter beats the SQL:
\begin{equation}\label{simplxi}
  \xi^2=\frac12\left[\frac{P_{simple}}{(\Omega\tau)^2}\cdot(1-r+2r(\Omega\tau)^2)+\frac{(\Omega\tau)^2}{P_{simple}}\frac{2(1+\ctg^2\Psi)}{(1-r)^2+4r(\Omega\tau)^2}+2\ctg\Psi\right]\,,
\end{equation}
where $P_{simple}=\dfrac{16\omega_\circ W\tau^2}{mc^2}$.

The main goal of optimization is to beat the SQL in the frequency
band that is as wide as possible. First of all we should find the
optimal value of $\ctg\Psi$ and $P_{simple}$. These parameters
should not depend on frequency therefore we will optimize them at
high frequencies, \textit{i. e.} $\Omega\tau\rightarrow\infty$.
This optimization will result in the following values:
\begin{equation}\label{simplPsi}
  \ctg\Psi=-2rP_{simple}\,,
\end{equation}
and being substituted to (\ref{simplxi}) will produce
\begin{equation}\label{simplxiHF}
  \xi^2_{\rm HF}\simeq\frac{1}{4rP_{simple}}\,.
\end{equation}

Obviously, the rise of $\xi^2$ at low frequencies is determined by
the radiation pressure spectral density as all other items in
(\ref{simplStotal}) are proportional to $\Omega^2$ and $\Omega^4$
and, therefore, can not influence at low frequencies. Hence we see
that optical losses lead to decrease of speed meter sensitivity
when observation frequency is small. It is useful to calculate the
value of parameter $P_{simple}$ that provides minimum of $\xi^2$
at defined frequency $\Omega^*$. It can be shown that for $1-r\ll1$ this value is equal to
\begin{equation}\label{simplePMin}
  P_{simple}^{opt}\simeq\dfrac{\Omega^*\tau}{\sqrt{2(1-r)}}\,.
\end{equation}
One can also readily obtain that minimal frequency where speed
meter sensitivity is equal to the SQL, \textit{i. e.} where
$\xi^2=1$ is fulfilled, is defined by the expression:
\begin{equation}\label{simpleOmegaMin}
  \Omega_{min}\simeq\dfrac{P_{simple}\sqrt{2(1-r)}}{\tau\sqrt{4P_{simple}-1}}\,,
\end{equation}
where $r$ is central mirror amplitude reflectivity coefficient
that characterizes scheme optical losses. Here we suppose $t\ll1$.

Comparison of expressions (\ref{simplxiHF}) and
(\ref{simpleOmegaMin}) shows that one needs to increase optical
power ($P_{simple}\rightarrow\infty$) in order to obtain high
sensitivity, while to have wide frequency band and increase the
sensitivity at low frequencies it is necessary to have low value
of pumping power to decrease radiation pressure noise. In Fig.
\ref{fig3} several curves are presented that demonstrates how
$\xi^2$ depends on frequency at different values of $P_{simple}$.
Chain lines are for ideal case where losses are equal to zero
($r=1$).

\section{Signal-recycled speed meter vs. signal-recycled position meter}\label{SRPMvsSRSM}
\begin{figure}[h]
\begin{center}

\psfrag{x1}{$x_1$}\psfrag{x2}{$x_2$}\psfrag{N1}{$N_{FP}$}\psfrag{N2}{$N_{SRM}$}\psfrag{ITM}{ITM}
\psfrag{SRM}{SRM}\psfrag{in}{in}\psfrag{out}[lb]{out}\psfrag{BS}{BS}
\includegraphics[height=60mm]{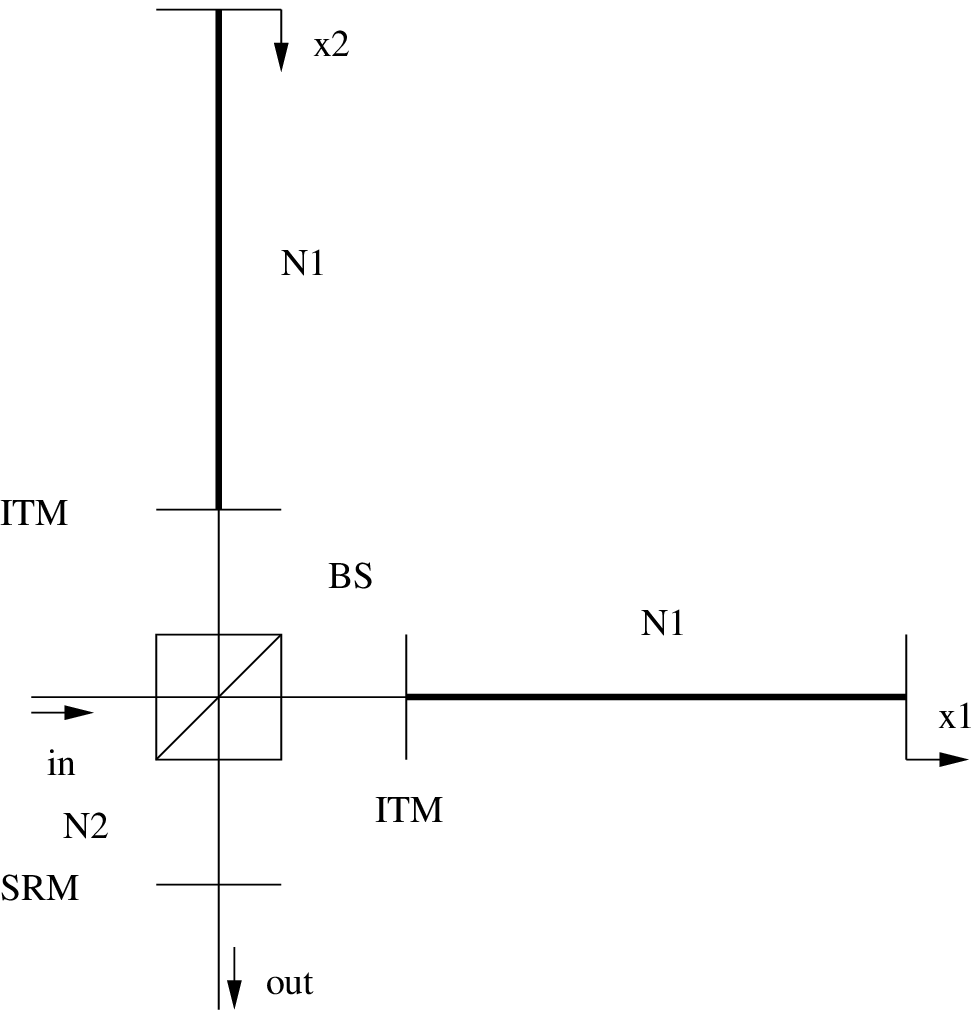}\hspace{25mm}\includegraphics[height=60mm]{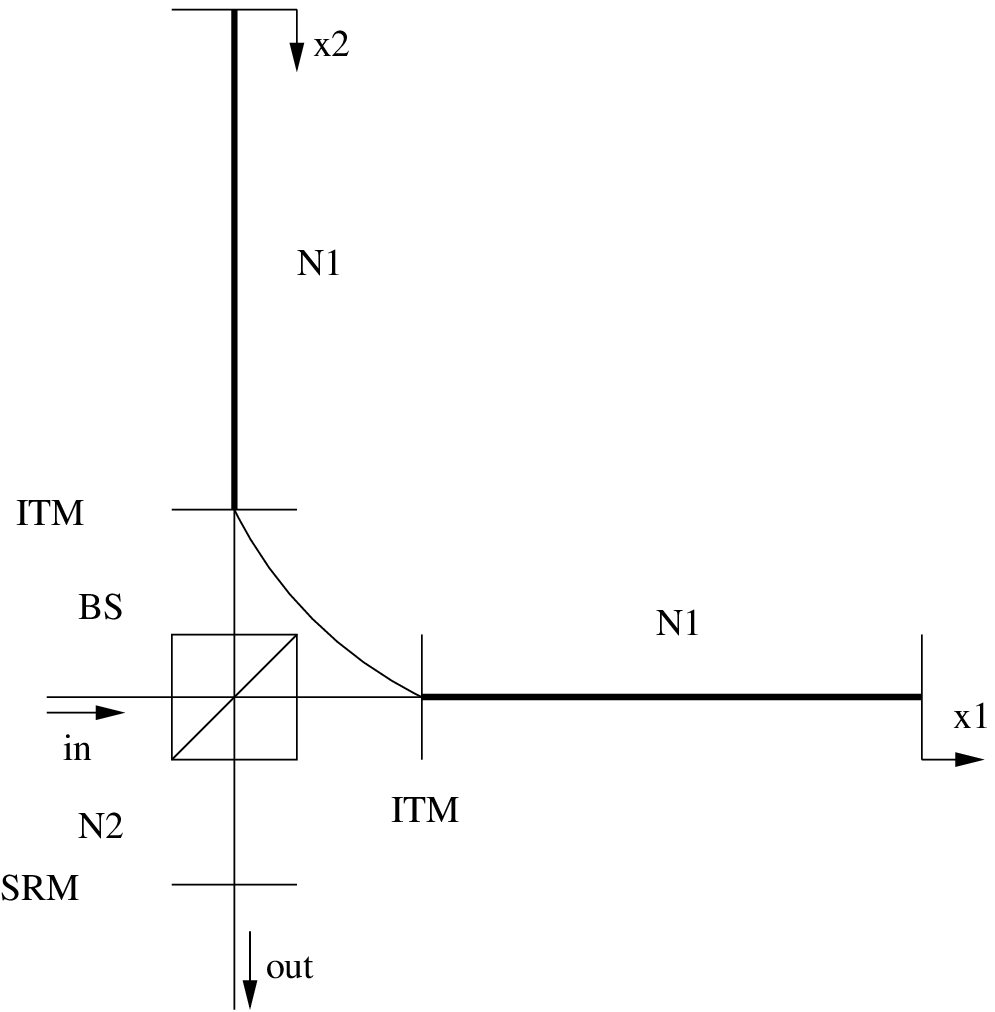}\\
\parbox[t]{0.48\textwidth}{\caption{Signal-recycled position meter}\label{figSRPos}}\hfill\parbox[t]{0.48\textwidth}{\caption{Signal-recycled speed meter}\label{figSRSM}}
\end{center}
\end{figure}

In this section we will consider semi-qualitatively the influence of signal recycling mirror on sensitivity of traditional position meter and speed meter. Let us consider the two schemes presented in Fig. \ref{figSRPos} and Fig. \ref{figSRSM}.
Suppose $T_{FP}\sim1/N_{FP}$ and $T_{SRM}\sim1/N_{SRM}$ are transmittances of arm cavities input mirrors (ITM) and signal-recycling mirror (SRM) correspondingly. Here $N_{FP}$ shows how many times light beam is reflected from the end mirrors until it leaves the cavity, and $N_{SRM}$ is the number of light beam reflections from the SRM before leaving  the whole scheme. The first number characterizes Fabry-Perot half-bandwidth, while the second one represents signal-recycling cavity half-bandwidth. Let $\tau$ be the time light needs to travel from one cavity mirror to another. Suppose also mirrors motion is uniform and can be expressed during light storage time ($\tau^* \sim N_{FP}N_{SRM}\tau < \tau_{GW} \sim \dfrac{\pi}{\Omega_{GW,\,max}}$) by formula
$$x_i=x_{i,\,0}+v_it\,,$$
where subscript $i=1,\,2$ denote the interferometer arm,
$x_{i,\,0}$ are the end mirrors initial position, and $v_i$ are
their velocities.

In the case of position meter light path is the following. After leaving the beam-splitter light enters both cavities, it passes from one arm cavity mirror to another  $2N_{FP}$ times, then it is reflected from the SRM and returns to the cavity. This cycle repeats $N_{SRM}$ times until the light leaves the scheme and get to the homodyne detector. Output beam phase shift in our simple case can be written as
\begin{equation}
\delta\varphi_{signal} \simeq \dfrac{2\omega_\circ}{c}N_{FP}N_{SRM}\bar x_-\,,
\end{equation}
where $\bar x_- = \bar x_1-\bar x_2$ is the mean end mirrors relative displacement during the storage time, and $c$ is the speed of light.

As for the speed meter, the light path  in this scheme differs from the previous one. Light that leaves, for example, the first cavity goes not to the beam-splitter and then to the SRM, but to the second cavity where it experiences $N_{FP}$ reflections from the end mirror in addition to those in the first cavity. Then it is reflected from the SRM and so on for $N_{SRM}$ times until it leaves the scheme. It can be shown that in this specific case the output beam phase shift will be written as
\begin{equation}
\delta\varphi_{signal} \simeq \dfrac{2\omega_\circ}{c}N_{SRM}N_{FP}^2\bar v_-\tau\,,
\end{equation}
where $\bar v_- = \bar v_1-\bar v_2$ is the relative velocity of the end mirrors.

We can readily see that position meter and speed meter output signals depend on arm cavities ($N_{FP}$) and signal recycling cavity ($N_{SRM}$) parameters in different way. It can be shown that this difference significantly influence the sensitivity of these schemes. Let us calculate the amount of circulating optical power necessary to achieve the level of SQL in each of the above mentioned schemes. The output signal field quadrature component $a_{signal}$ for both schemes is proportional to the product 
$$a_{signal}\simeq A\delta\varphi_{signal}\,,$$
where $A$ is the input radiation field quadrature amplitude. Therefore spectral density of phase fluctuations can be expressed in terms of spectral density of $a_{signal}$  as 
$$S_{\varphi}=\dfrac{S_a}{|A|^2}\,.$$
The expression for $S_a$ can be readily obtained and will be equal to 
$$S_a=\dfrac14\,,$$
and $|A|^2$ can be represented in terms of power circulating in arms as 
$$|A|^2=\dfrac{W}{\hbar\omega_\circ N_{FP}N_{SRM}}\,.$$
Then for $S_\varphi$ one will have the following expression
\begin{equation}\label{SphiGen}
S_\varphi=\dfrac{\hbar\omega_\circ N_{FP}N_{SRM}}{4W}\,.
\end{equation}
On the other hand, for position meter this spectral density can be expressed in terms of caused by radiation shot noise mirror displacement spectral density corresponding to the SQL $S_x^{SQL}=\frac{\hbar}{m\Omega_\circ^2}$ as
\begin{equation}\label{SphiPM}
S_\varphi^{PM}=\dfrac{4\omega_\circ^2N_{FP}^2N_{SRM}^2}{c^2}S_x^{SQL}=\dfrac{4\hbar\omega_\circ^2N_{FP}^2N_{SRM}^2}{mc^2\Omega_\circ^2}\,,
\end{equation}
where $\Omega_\circ$ is some fixed observation frequency, $m$ is the mirror mass, and for speed meter in terms of caused by radiation shot noise mirror velocity spectral density corresponding to the SQL $S_v^{SQL}=\frac{\hbar}{m}$ as
\begin{equation}\label{SphiSM}
S_\varphi^{SM}=\dfrac{4\omega_\circ^2\tau^2N_{FP}^4N_{SRM}^2}{c^2}S_v^{SQL}=\dfrac{4\hbar\omega_\circ^2\tau^2N_{FP}^4N_{SRM}^2}{mc^2}\,.
\end{equation}

Substituting (\ref{SphiGen}) into (\ref{SphiPM}) and (\ref{SphiSM}) we will obtain the following values of circulating powers for both schemes considered:
\begin{equation}
W_{PM}=\dfrac{mc^2\Omega_\circ^2}{16N_{FP}N_{SRM}\omega_\circ}\,,\quad W_{SM}=\dfrac{mc^2}{16N_{FP}^3N_{SRM}\omega_\circ\tau^2}\,.
\end{equation}
If we divide $W_{SM}$ by $W_{PM}$ we will obtain the factor of $\chi$ by which speed meter circulating power necessary to achieve the SQL is larger than one for the position meter
\begin{equation}
\chi=\dfrac{W_{SM}}{W_{PM}}=\dfrac{1}{(N_{FP}\Omega_\circ\tau)^2}\gg N_{SRM}^2\,,
\end{equation}
where we have taken into account that light storage time should be much smaller than period of gravitational wave, \textit{i. e.} $\tau^*\sim N_{FP}N_{SRM}\tau\ll\pi/\Omega_\circ$. Obviously, the above estimates show that speed meter with signal recycling hardly can be considered as the best variant for implementation as QND-meter, as it requires much more circulating power than signal-recycled position meter with the same parameters. More precise calculations that prove the above simple considerations are performed in Appendix \ref{App2a}.

\section{Optical losses in speed meter interferometers with Fabry-Perot cavities in arms.}\label{Sec3}
\subsection{Speed meter interferometer action and output signal}\label{SubSec31}

Let us consider the scheme of gravitational-wave detector based on  speed meter
principle presented in Fig. \ref{fig4}.
\begin{figure*}[t]\begin{center}

\begin{picture}(120,120)
  \thinlines\drawline(0,15)(45,15)(45,110)\drawline(15,0)(15,35)(110,35)
  \put(0,0){\usebox{\Speedmeter}}
\end{picture}

\caption{Speed meter interferometer} \label{fig4}

\end{center}\end{figure*}

This interferometer differs from the traditional LIGO
interferometers by additional polarization beam-splitter (PBS) and
two quarter-wave plates ($\lambda/4$). Quarter-wave plates are
needed to transform input light polarization from linear
($0^\circ$ and $90^\circ$) to circular one ($\circlearrowleft$ and
$\circlearrowright$) during each pass.  After reflecting from one
of 4 km long Fabry-Perot (FP) cavities light polarization varies
over $90^\circ$, and light beam is reflected from the PBS to the
second FP cavity. Hence, the light beam passes not only one FP
cavity, but both cavities consequently. As there are two beams
that propagate in opposite directions (in clockwise and
counter-clockwise directions correspondingly) the scheme output
does not contain any information about the symmetric mechanical
mode of the mirrors, \textit{i. e.}  about the sum of end mirrors
displacements. This mode is not coupled with gravitational-wave
signal and its presence can significantly lower the sensitivity,
and prevent from beating the SQL. The output signal of this scheme
can be found in the same manner as it was done in \cite{02a2Kh}.
The only problem is to include additional noises, arising due to
internal losses in optical elements to calculations. To do so,
suppose each mirror to be a source of two independent additional
noises, that stand for interaction of electromagnetic radiation
with mirror medium excitations. This approach is based on the
Huttner-Barnett scheme \cite{92a1HuBa} of electromagnetic field
quantization in linear lossy dielectrics. Using this approach one
can obtain input-output relations for any optical device with
losses (see \cite{95a1GrWe,99a1KnSchSChWeCh}).

Let us use the same notations as in Section \ref{Sec2} for the
scheme parameters. In order to distinguish values that correspond
to inner and end mirror we will denote the first ones by
subscript $_1$ and the second ones by subscript $_2$. Each
Fabry-Perot has the following parameters: mirrors reflectivity,
transmittance, and absorption coefficients are equal to
$-r_1,\,-r_2$, $it_1,\,0$, and $ia_1,\,ia_2$ correspondingly, and
cavity length is $L$. The end mirrors transmittance we suppose to
be equal to zero because there is no difference whether light run
out or is absorbed in the end mirror.

Suppose that classical pumping wave amplitude is $A$ and sideband
fluctuations operator is $\hat a_w$. Suppose also zero
oscillations that enter the scheme from the "southern" side are
described by operator $\hat a_{s}$. We will also need operators
$\hat g_{s_{11}}$, $\hat g_{s_{12}}$, $\hat g_{s_{21}}$, and $\hat
g_{s_{22}}$ to describe noises due to internal losses in FP
mirrors (the second numerical subscript stands for the number of
noise that arises in the mirror due to internal losses).
Superscripts $^I$ and $^{II}$ denote what noises arise during the
first and the second reflection correspondingly. Parameter $\pmb{\eta}$ we assume to be equal to unity, where $\pmb{\eta}$ has the same meaning as in simple scheme and is equal to the ratio of classical amplitudes inside the cavity during the
second and the first reflection correspondingly (see Appendix
\ref{App2}).

Finally, we are able to write down the output sideband
fluctuations operator $\hat c_s$ that contains information about the mirrors movements, and therefore about the
gravitational wave force:
\begin{multline}\label{SMout}
  \hat c_{s}= \frac{1}{{\cal L}^2(\omega)}[i{\cal
B}^2_1(\omega)\hat a_{s}^I(\omega)-e^{2i\omega\tau}r_2t_1a_1\hat
g_{s_{11}}^I(\omega)+e^{i\omega\tau}t_1a_2\hat
g_{s_{21}}^{I}(\omega))-\\-i{\cal
L}(\omega)(e^{2i\omega\tau}r_2t_1a_1\hat
g_{w_{11}}^{II}(\omega)-e^{i\omega\tau}t_1a_2\hat
g_{w_{21}}^{II}(\omega))]-{\cal K}_{SM}x_-(\omega_\circ-\omega)\,.
\end{multline}
where ${\cal L}(\omega)=r_1r_2e^{2i\omega\tau}-1$, ${\cal
B}_{1}(\omega)=r_1-e^{2i\omega\tau}r_2(r_1^2+t_1^2)$, and $${\cal
K}_{SM}=-\dfrac{({\cal B}_1(\omega)-{\cal
L}(\omega))E\kappa(\omega)e^{i\omega\tau}r_2t_1}{{\cal
L}^2(\omega)}$$ is the coupling constant of the speed meter
interferometer, $E$ is the classical amplitude of pumping field
near the movable mirror after the first reflection.

\subsection{Speed meter spectral densities and sensitivity}

Here we will obtain the expressions for radiation pressure noise, shot noise and their cross-correlation spectral
densities. We will suppose that the same additional
pumping procedure as in Section \ref{Sec2} is taking place.
Here we will use the same approximation that are defined by
(\ref{NBconditions1}) of Section \ref{Sec2}. It is useful to
evaluate if this approximation is valid for LIGO. Parameters for
LIGO interferometer are the following:
\begin{equation*}
  \omega_\circ=1.77\cdot10^{15}\,\mbox{s}^{-1},\quad
  \tau=1.33\cdot10^{-5}\,\mbox{s}\,,\quad m=40\mbox{ kg}\,,\quad L=4\cdot10^3\mbox{ m}\,.
\end{equation*}
If we suppose that gravitational signal upper frequency is about $\Omega=10^3$ s$^{-1}$ then there remain no doubts that for LIGO condition of narrow-band approximation applicability is fulfilled as $\Omega\tau=1.33\cdot10^{-2}\ll 1$.

It seems reasonable to introduce the following notations:
\begin{subequations}\label{NBnotations2}
\renewcommand{\theequation}{\theparentequation.\arabic{equation}}
\begin{equation}
   \gamma_1=\frac{t_1^2}{4\tau}\,,
\end{equation}
\begin{equation}
  \alpha=\alpha_1+\alpha_2=\frac{a_1^2+a_2^2}{4\tau}\,,
\end{equation}
\begin{equation}
  \gamma=\frac{1-r_1r_2}{2\tau}=\gamma_1+\alpha\,.
\end{equation}
\end{subequations}
Using this notations one can obtain:
\begin{subequations}
\renewcommand{\theequation}{\theparentequation.\arabic{equation}}
\begin{equation}
  {\cal B}_1(\omega_\circ -\Omega)=2\tau(\gamma-2\gamma_1+i\Omega)\,,
\end{equation}
\begin{equation}
   {\cal L}(\omega_\circ -\Omega)=-2\tau(\gamma+i\Omega)\,.
\end{equation}
\end{subequations}

Taking all of the above into account one will obtain the following
expressions for spectral densities (see derivation in Appendix
\ref{App2})
\begin{subequations}\label{SMspdens}
\renewcommand{\theequation}{\theparentequation.\arabic{equation}}
\begin{equation}\label{SMSx1}
  S_x(\Omega)=\frac{\hbar
  L^2}{32\omega_\circ \tau W}\frac{(\gamma^2+\Omega^2)^2}{2\gamma_1\sin^2\Psi((\gamma-\gamma_1)^2+\Omega^2)}\,,
\end{equation}
\begin{equation}\label{SMSf1}
  S_{F}(\Omega)=\frac{8\hbar\omega_\circ\tau
  W}{L^2}\frac{\gamma(\gamma^2+\Omega^2)-\gamma_1(\gamma^2-\Omega^2)}{(\gamma^2+\Omega^2)^2}\,,
\end{equation}
\begin{equation}\label{SMSxf1}
  S_{xF}=-\dfrac{\hbar}{2}\cot\Psi\,,
\end{equation}
\end{subequations}
Where $\Psi$ is the homodyne angle, and $W$ is the pumping power at the end mirrors.

One can obtain that the minimum force that can be measured by
speed meter presented in Fig. \ref{fig4} depends on the total
measurement noise spectral density. This spectral density can be
expressed by the formula (\ref{simplStotal}) but one exception:
$m$ in this formula is equal to one fourth of the real end mirror
mass. The multiplier $1/4$ appears when we suppose not only end
mirrors to be movable, but also the inner ones.

\begin{figure}[h]
\begin{center}
\psfragscanon \psfrag{W}[lt]{$\upsilon=\Omega/\gamma$}\psfrag{xi}[cb]{$\xi$}
\includegraphics[width=0.48\textwidth]{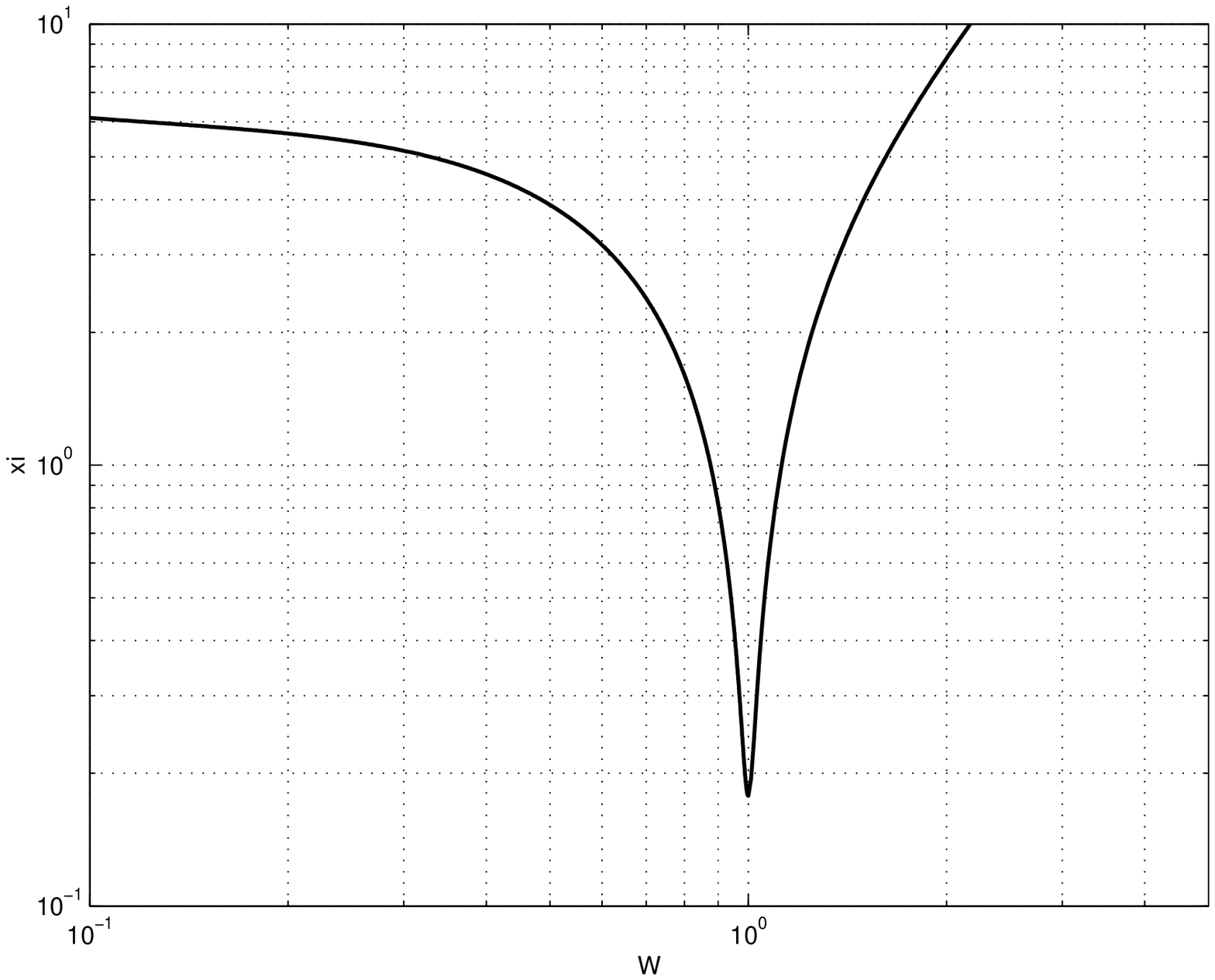}\hfill\includegraphics[width=0.48\textwidth]{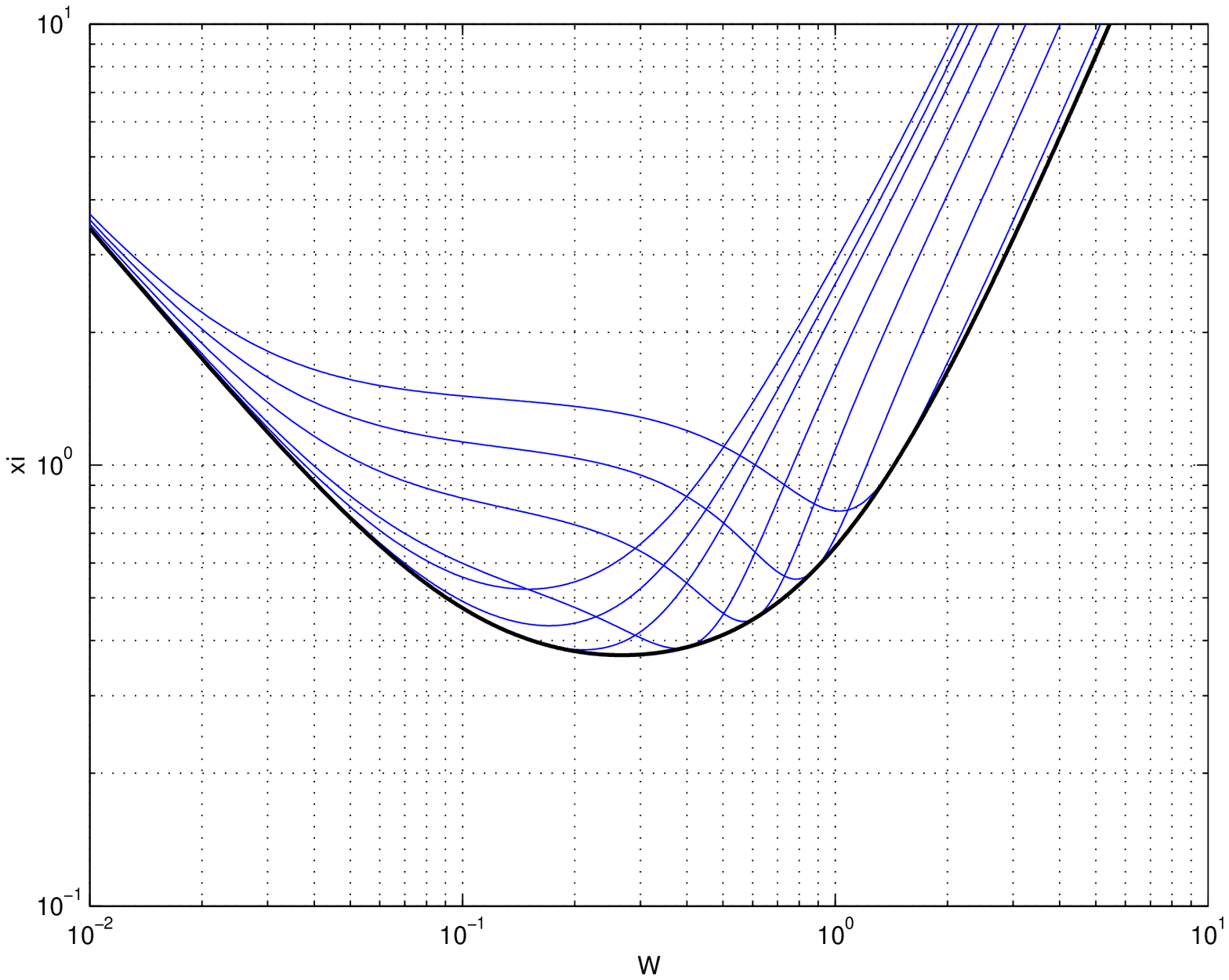}\\
\parbox[t]{0.48\textwidth}{\caption{Plot of $\xi=\sqrt{S_{SM}/S_{SQL}}$, optimized at some fixed frequency $\Omega_\circ$. For given $\Omega_\circ=10^3$ s$^{-1}$ and fixed optical losses contribution to the total half-bandwidth $\alpha=1$ s$^{-1}$ one should take the following parameters to obtain the best sensitivity at given frequency: total half-bandwidth $\gamma=\Omega_\circ$, circulating power $W=\frac{mL^2\Omega_\circ^3}{8\omega_\circ\tau}\sqrt{\frac{\Omega_\circ}{\alpha}}\simeq27$ MW, and $\cot\Psi=-\sqrt{\frac{\Omega_\circ}{\alpha}}\simeq-3\cdot10^{-2}$. The minimal $\xi$ in that case will be equal to $\sqrt[4]{\frac{\alpha}{\Omega_\circ}}\simeq0.18$.}\label{fig5}}\hfill\parbox[t]{0.48\textwidth}{\caption{Plots of $\xi$ for circulating optical power $W=1$ MW and several fixed homodyne angles (thin blue curves; the curve with rightmost minimum corresponds to the least angle). Bold black curve corresponds to frequency-dependant homodyne angle.}\label{fig5a}}\\
\end{center}
\end{figure}
\begin{figure}[h]
\begin{center}
\psfragscanon \psfrag{W}[lt]{$\upsilon=\Omega/\gamma$}\psfrag{xi}[cb]{$\xi$}
\includegraphics[width=0.48\textwidth]{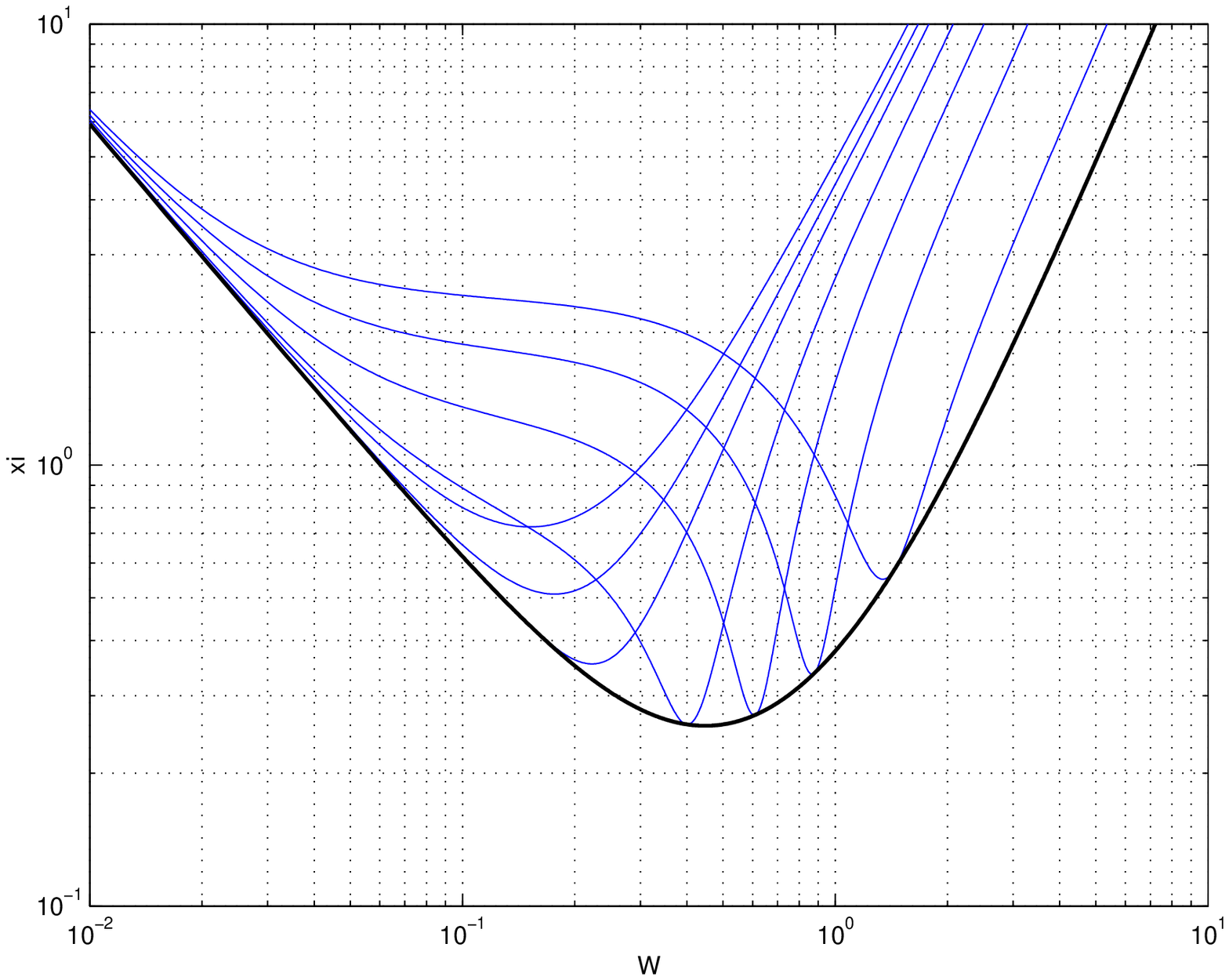}\hfill\includegraphics[width=0.48\textwidth]{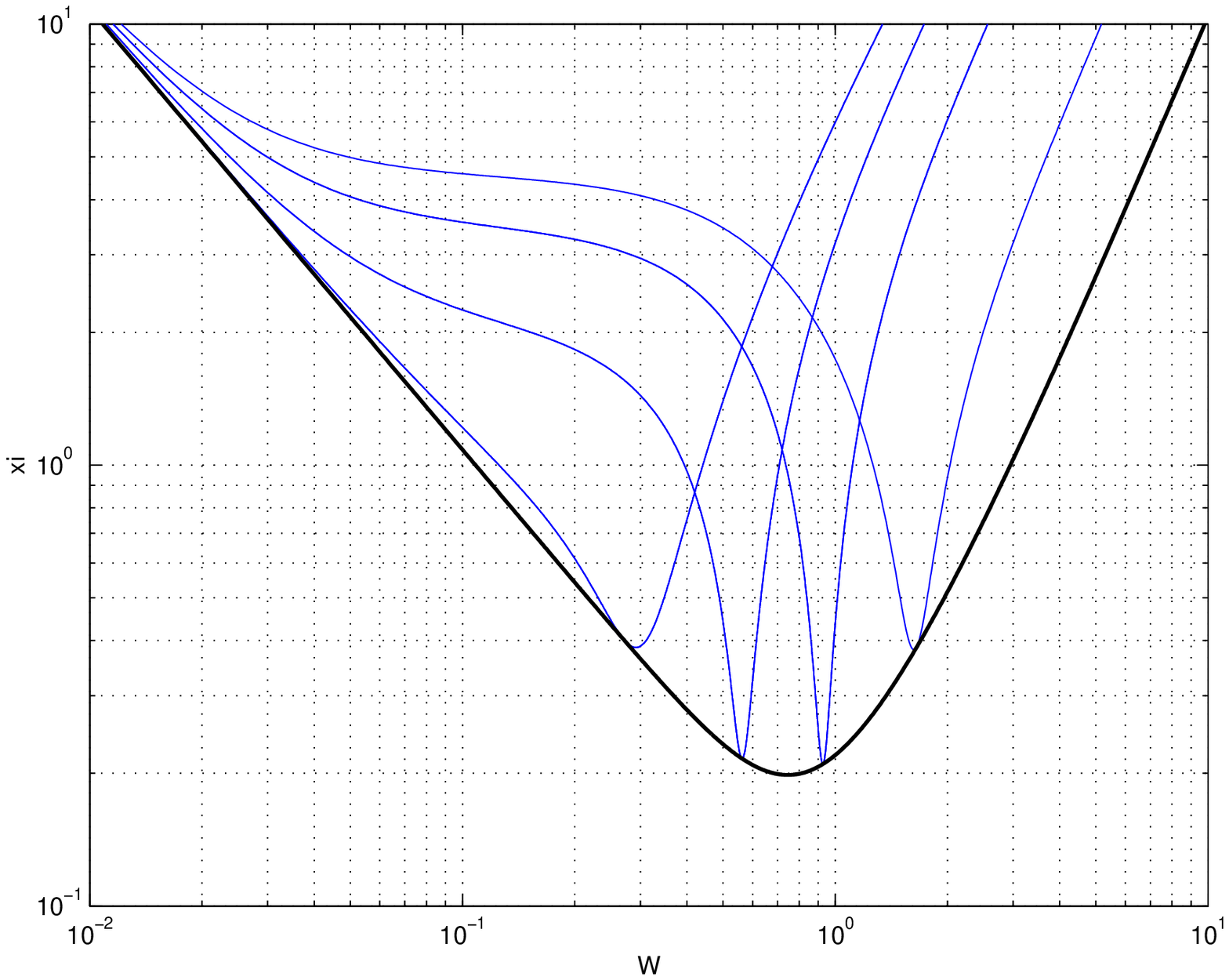}\\
\parbox[t]{0.48\textwidth}{\caption{Plots of $\xi$ for circulating optical power $W=3$ MW and several fixed homodyne angles (thin blue curves; the curve with rightmost minimum corresponds to the least angle). Bold black curve corresponds to frequency-dependant homodyne angle.}\label{fig5b}}\hfill\parbox[t]{0.48\textwidth}{\caption{Plots of $\xi$ for circulating optical power $W=10$ MW and several fixed homodyne angles (thin blue curves; the curve with rightmost minimum corresponds to the least angle). Bold black curve corresponds to frequency-dependant homodyne angle.}\label{fig5c}}
\end{center}
\end{figure}

Now we can write down the expression for $\xi^2=S_{\rm SM}/S_{\rm
SQL}$, where $S_{\rm SM}$ is the total quantum noise of the speed
meter that is calculated in accordance with (\ref{simplStotal}),
and $S_{\rm SQL}=\hbar m\Omega^2$ is the SQL spectral density for
fluctuational force:
\begin{equation}\label{SMxi}
  \xi^2=\frac{8\omega_\circ\tau
  W}{mL^2\Omega^2}\frac{\gamma(\gamma^2+\Omega^2)-\gamma_1(\gamma^2-\Omega^2)}{(\gamma^2+\Omega^2)^2}+\frac{m
  L^2\Omega^2}{32\omega_\circ \tau
  W}\frac{(\gamma^2+\Omega^2)^2}{2\gamma_1\sin^2\Psi((\gamma-\gamma_1)^2+\Omega^2)}+\cot\Psi\,.
\end{equation}
This value characterizes scheme sensitivity. Our goal is to have
this value as small as possible. In the next subsection we will perform two possible optimizations of $\xi^2$, correspondingly in narrow and wide frequency bands.

\subsection{Optimization of $\xi^2$}
\paragraph{Narrow-band optimization.}

Let optimize (\ref{SMxi}) at some fixed frequency. The possible situation when such optimization can be useful is the detection of gravitational radiation emitted by quasi-monochromatic sources. Compact quikly rotating neutron stars, \textit{i. e.} pulsars, may be the example of such sources. Gravitational radiation from pulsars is quasy-monochromatic and relatively weak so it is crucial to have high sensitivity in narrow frequency band to detect these sources. So it is convenient to find the minimum of $\xi^2$ at the source main frequency $\Omega_\circ$. If we suppose that optical losses are small enough, \textit{i. e.}  $\alpha\ll\Omega_0$ then the minimal value of $\xi$ will be equal to
\begin{equation}\label{xi_min}
\xi_{min}=\sqrt[4]{\varepsilon}=\sqrt[4]{\dfrac{\alpha}{\Omega_\circ}}\,,
\end{equation} 
and is reached at $\gamma_{opt}=\Omega_\circ$, optical circulating power $W_{opt}=\dfrac{mL^2\Omega_\circ^3}{8\omega_\circ\tau}\sqrt{\frac{\Omega_\circ}{\alpha}}$, homodyne angle defined by formula $\cot\Psi_{opt}=-\sqrt{\dfrac{\Omega_\circ}{\alpha}}$ (see Appendix \ref{App3} for detail). For $\alpha=1$ s$^{-1}$ it will be equal to $\xi\simeq0.18$. For LIGO interferometer optical power necessary to reach the above value of $\xi$ is equal to $W_{opt}\simeq27$ MW.

Unfortunately, at frequencies different from $\Omega_\circ$ the value of $\xi$ is much worse. Function $\xi$ behaviour at different frequencies is presented in Fig. \ref{fig5}. We can see that significant gain in sensitivity compared to the SQL can be achieved in a very narrow frequency band of about several tens of hertz. Thus, this regime of speed meter operation that we prefer to call "narrow-band" regime can be used to beat the SQL by significant amount only in narrow band near some arbitrarily chosen frequency for the purposes of weak quasy-monochromatic sources detection.

\paragraph{Wide-band optimization.}

Contrary to the narrow-band case, considered above, the vast majority of gravitational wave sources either radiate in relatively wide frequency band, or their main frequency is unknown. In both cases it is necessary to perform wide-band detection procedure. This problem can not be solved as easy as the previous one, because there are no criteria what should be the frequency bandwidth value and what sensitivity should be reached within the ranges of this band. As there is no clarity in this question it seems convenient to represent the variety of different possible regimes of operation that the speed meter is capable of. Therefore we will suppose that optical power $W$ circulating in the arms of interferometer and resonators half-bandwidth $\gamma$ are fixed.   Then varying homodyne angle $\Psi$ it is possible to obtain variety of different sensitivity curves. These curves for three different values of circulating optical power $W=1,\,3$, and $10$ MW are presented by thin blue curves in Figures \ref{fig5a}, \ref{fig5b}, and \ref{fig5c} correspondingly. Varying homodyne angle it is possible to reach sensitivity even three times better than SQL in relatively wide frequency band. It is also possible to choose the frequency where the best sensitivity is reached by changing the homodyne angle, \textit{i. e.} increase of $\Psi$ leads to sensitivity curve offset into the lower frequencies domain. It should be also noted that the increase of circulating power in this regime leads to reduction of frequency band where $\xi$ is less than $1$, therefore it is reasonable to use the moderate values of optical powers.

Using the technique of frequency-dependant variational readout suggested in article \cite{02a1KiLeMaThVy} it is possible to increase the sensitivity of speed meter significantly in wide frequency band. The achievable sensitivities in this case are represented in the same figures by bold black curves. Here is the dependence of $\Psi(\Omega)$ that allows to obtain the above mentioned results:
\begin{equation}
\Psi(\Omega)=\pi-\arctan\left(\dfrac{mL^2}{32\omega_\circ\tau W}\cdot\dfrac{\Omega^2(\gamma^2+\Omega^2)^2}{\gamma_1(\alpha^2+\Omega^2)} \right)\,. 
\end{equation}

\section{Conclusion}\label{Sec4}
In this section we will sum up the results of our consideration.
The following conclusions can be made:
\begin{itemize}
 \item The attempts to increase speed-meter sensitivity by introducing signal-recycling mirror (SRM) or replace arm cavities by this mirror will not be successful, because contrary to the position meter, speed meter sensitivity depends upon the transmittances of arm cavities input mirrors (ITM) and SRM not symmetrically. The influence of ITM is considerably greater than SRM influence because in speed meter light beam passes consequently through both cavities before it is reflected from SRM. Therefore, it is convenient to use arm cavities in gravitational-wave detectors based on speed-meter principle.
  \item Speed meter topology of gravitational-wave antennae allows to achieve sensitivity about three times better than SQL in relatively wide frequency band preserving optical power circulating in the arms at reasonable level of $1$ MW. Moreover, its sensitivity can be improved and its frequency band can be significantly increased if one applies variational readout technique. 
  \item  The ultimate sensitivity speed meter is capable of is defined by two factors, optical losses and bandwidth  of arm cavities and, therefore, circulating power. This sensitivity can be expressed in terms of these factors as
$$\xi=\dfrac{h}{h_{SQL}}=\sqrt[4]{\dfrac{\alpha}{\gamma}}\,,$$
where $h$ is the metric variation that can be measured by speed meter, $h_{SQL}$ is the standard quantum limit for $h$, $\alpha$ is the optical losses contribution to the total interferometer half-bandwidth $\gamma$. However, high sensitivity (small $\xi$) requires large amount of circulating optical power (about tens of megawatts) and can be achieved in relatively narrow frequency band. In our opinion, the best operation mode for speed meter is the wide-band regime with frequency dependant readout.

\end{itemize}

\section{Acknowledgements}\label{Sec5}
Author would like to express his sincere gratitude to professor F.
Ya. Khalili for interesting subject of investigation and help and
advice. Author also would like to thank V. B. Braginsky, S. P.
Vyatchanin, and S. E. Strigin for helpful discussions and
countenance. This paper is supported in part by NSF and Caltech
grant \#PHY0098715, by the Russian Foundation for Basic Research,
and by Russian Ministry of Industry and Science.
\appendix

\section{Derivation of spectral densities for simple case}\label{App1}
\subsection{Input-output relations derivation}
Here we will obtain the formulae for quantum noise spectral
densities, presented in Section \ref{Sec2} by expressions
(\ref{simpleSpDens}). The input light can be described by the
formula (\ref{EMFquant}). Input amplitudes are the following
(Subscripts "w, e, s, n" are for light beams corresponding to the
consequent part of the scheme. For example, $A_w$ means classical
amplitude of light beam propagating in "western" direction.):
\begin{equation}
  A_w=A\,,\quad A_s=0\,,
\end{equation}
and corresponding sideband operators are
\begin{equation}
  \hat a_w\,,\quad \hat a_s\,.
\end{equation}
As we have mentioned above we are able to introduce additional
pumping through the central mirror. This additional pumping
increases light power in the scheme and does not create additional
noises, therefore, it increases scheme sensitivity. To describe it
we will introduce complex parameter $\pmb{\eta}= \eta
e^{i\Phi}=\sqrt2C/A$ that is equal to the ratio of light amplitude
at the end mirrors with additional pumping($C=A/\sqrt2+A_{add}$)
and light amplitude after the beam splitter ($A/\sqrt2$) (that is
equal to the amplitude at the end mirror provided that central
mirror is ideally reflecting, \textit{i. e.} in lossless case).

Taking all above into account one can easily obtain that classical
output of the scheme is equal to

\begin{equation}
  B_s=0\,,
\end{equation}
and corresponding sideband output is equal to
\begin{multline}
  \hat b_s(\omega)=ir\hat a_s(\omega)e^{4i\omega\tau}-i\alpha\frac{-\hat g_n(\omega)+i\hat g_e(\omega)}{\sqrt2}e^{2i\omega\tau}-\\-2\kappa(\omega)A(ire^{i(\omega_\circ +3\omega)\tau}-\pmb{\eta} e^{i(\omega_\circ +\omega)\tau})x_-(\omega_\circ -\omega)=\beta_{input}\hat a_s + \beta_{loss}\hat g_s+{\cal
  K}_{simple}x_-\,,
\end{multline}
see (\ref{SimpleOutput}) for notations.

\subsection{Radiation pressure noise spectral density $S_F(\Omega)$}\label{SimpleSMSf}
The radiation pressure force corresponding to system mode $x_-$ is
equal to:
\begin{multline}\label{Ffluct}
  \hat F = \hat F_e-\hat F_n =  2\hbar\int_0^\infty\kappa(\omega)A^*[(ie^{i(\omega-\omega_\circ )\tau}+\eta e^{-i\Phi}re^{i(3\omega-\omega_\circ )\tau})\hat a_s(\omega)-\\-\sqrt2\alpha \eta e^{-i\Phi}e^{i(\omega-\omega_\circ )\tau}\frac{-\hat g_n(\omega)+i\hat g_e(\omega)}{\sqrt2}]e^{i(\omega_\circ -\omega)t}\frac{d\omega}{2\pi}+\rm
  h.\,c.\,,
\end{multline}
where $\hat F_e$ and $\hat F_n$ are the radiation pressure
fluctuational forces acting upon "eastern" and "northern" movable
mirrors correspondingly.

In order to calculate radiation pressure spectral density one
should calculate symmetric correlation function:
\begin{equation}
  B_F(t-t')=\frac12\langle0|(\hat F(t)\hat F(t')+\hat F(t')\hat
  F(t))|0\rangle\,,
\end{equation}
where $|0\rangle$ --- is the radiation field ground state. If we
calculate this value we will obtain:
\begin{equation}
  B_F(t-t')=\frac12\left\{\int_0^\infty|F(\omega)|^2e^{i(\omega_\circ -\omega)(t-t')}\frac{d\omega}{2\pi}+\int_0^\infty|F(\omega)|^2e^{-i(\omega_\circ -\omega)(t-t')}\frac{d\omega}{2\pi}\right\}
\end{equation}
where
$$|F(\omega)|^2=4\hbar^2\kappa^2(\omega)|A|^2(|(e^{i(\omega-\omega_\circ
)\tau}-i\eta e^{i\Phi} re^{i(3\omega-\omega_\circ
)\tau})|^2+\alpha^2 \eta^2 )\,.$$ In order to obtain this result
one should take into account that

$\langle0|\hat a(\omega)\hat
a^\dagger(\omega')|0\rangle=2\pi\delta(\omega-\omega')$. The
radiation pressure spectral density can be defined as:
\begin{equation}\label{Sfdef}
  S_F(\Omega)=\int_{-\infty}^\infty B_F(t)e^{-i\Omega
  t}dt=\frac12(S'(\Omega)+S'(-\Omega))=\frac{16\hbar\omega_\circ  W}{c^2}\cdot\dfrac{1/2+\eta^2/2-\eta
  r\sin\Phi\cos2\Omega\tau}{1+\eta^2}\,.
\end{equation}
In order to provide speed meter mode of operation we need to set
$\Phi=\pi/2$, then one will obtain that
\begin{equation}
 S_F(\Omega)=\dfrac{8\hbar\omega_\circ W}{c^2}\cdot\dfrac{1+\eta^2-2\eta
  r\cos(2\Omega\tau)}{1+\eta^2}\,.
\end{equation}

\subsection{Shot noise spectral density $S_x(\Omega)$}\label{SimpleSMSx}
The output signal of the scheme is mixed up with local oscillator
wave in order detect phase shift due to end mirrors displacement
in the optimal way. This mixed radiation enters the homodyne
detector. The photo-current of detector is proportional to the following time-averaged value:
\begin{multline}\label{SMOutSign}
  \hat I_{p.d.}(t)\sim\overline{2E_{b_s}(t)\cos(\omega_\circ  t+\phi_{LO})}=\int_0^\infty\sqrt\omega\hat b_s(\omega)e^{i(\omega_\circ -\omega)t+i\phi_{LO}}\frac{d\omega}{2\pi}+\rm
  h.\,c.=\\=\int_{-\infty}^\infty K(\Omega)(\hat x_{fluct}(\Omega)+x_-(\Omega))e^{i\Omega t}\frac{d\Omega}{2\pi}\,,
\end{multline}
where $\phi_{LO}$ is the local oscillator phase, $\hat r(\Omega)$ is the noise operator which spectral
density is equal to unity, and $K(\Omega)$ is equal to:
\begin{equation}
  K(\Omega) = -\frac{4\omega_\circ ^{3/2}|A|}{c}e^{-3i\Omega\tau}[r\sin\Psi+\eta
  e^{2i\Omega\tau}\cos(\Phi+\Psi)]\,,
\end{equation}
where $\Psi=\phi_{\rm LO}+\arg A$. Fluctuations of coordinate are
described by the operator
\begin{equation}\label{xfluct}
  \hat x_{fluct}(\Omega)=\dfrac{\hat r(\Omega)}{K(\Omega)}\,.
\end{equation}
At last, assuming $\Phi=\pi/2$, the coordinate noise spectral
density is equal to
\begin{equation}
  S_x(\Omega)=\dfrac{\hbar c^2}{32\omega_\circ
 W\sin^2\Psi}\cdot\dfrac{1+\eta^2}{r^2+\eta^2-2r\eta\cos(2\Omega\tau)}\,,
\end{equation}

\subsection{Cross-correlation spectral density $S_{xF}(\Omega)$}

Now we know everything to calculate the cross-correlation spectral
density $S_{xF}(\Omega)$. In order to do it one should find the
cross-correlation function of (\ref{Ffluct}) and $\hat
x_{fluct}(\Omega)$. Using the same algorithm as in previous
subsections one will obtain that
\begin{equation}\label{simplSxf}
  S_{xF}(\Omega)=-\frac\hbar2\cot\Psi\,.
\end{equation}

 \section{Power- and signal-recycled speed meter
interferometer sensitivity. Precise analysis}\label{App2a}

\subsection{Input-output relations for power- and signal-recycled speed-meter interferometer.}
In this appendix we will analyze the scheme of speed meter
interferometer with power (PRM) and signal (SRM) recycling mirrors installed
(See Fig. \ref{fig3a}). Here we will confirm by exact calculations the results of qualitative consideration presented in Section \ref{SRPMvsSRSM}.
\setcounter{paragraph}{0}

\begin{wrapfigure}[14]{r}{0.33\textwidth}
\psfrag{L}[lb]{$L$}\psfrag{A1}[lb]{$a'$}\psfrag{A2}[lb]{$a''$}\psfrag{A3}[lb]{$a$}\psfrag{B1}[lb]{$b'$}\psfrag{B2}[lb]{$b''$}\psfrag{B3}[lb]{$b$}\psfrag{PRM}[lt]{PRM/SRM}
    \includegraphics[width=0.33\textwidth]{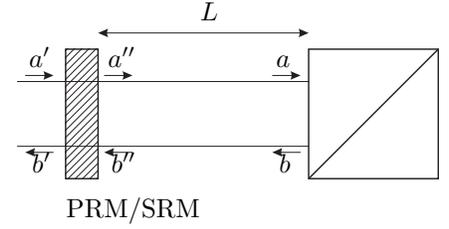}\\
  \caption{Power recycling/signal recycling mirrors}\label{figPR}
\end{wrapfigure}

Let consider the influence of additional optical elements on the
input-output relations.

Figure \ref{figPR} represents the situation common for PRM and
SRM. Beams $a$ and $b$ represent light entering and leaving the
beam-splitter, and beams $a'$ and  $b'$ stand for light, entering and leaving the PRM/SRM. The length $L$ of recycling cavity should be chosen in the way that provides the maximal values of circulating power in case of PRM, and signal sidebands in case of SRM.

\paragraph{Power recycling mirror.}
Power recycling influences the value of circulating power only, as quantum fluctuations from the laser that are influenced by the PRM, return back to the laser due to dark port tuning of interferometer and do not contribute to the output signal. Interferometer circulating power depends upon the quadrature amplitude $A$ of the beam $a$ as $$W=\hbar\omega_\circ|A|^2\,.$$

\begin{wrapfigure}[25]{l}{0.48\textwidth}
\begin{center}

\psfrag{A1}{$a_w$}\psfrag{As}{$a_s$}\psfrag{An}{$a_n$}\psfrag{Ae}{$a_e$}
\psfrag{Bw}{$b_w$}\psfrag{Bs}{$b_s$}\psfrag{Bn}[lb]{$b_e$}\psfrag{Be}{$b_n$}
\psfrag{Gn}{$g_n$}\psfrag{Ge}{$g_e$}\psfrag{Xn}{$x_n$}\psfrag{Xe}{$x_e$}\psfrag{PRM}[lb]{PRM}\psfrag{SRM}{SRM}
\includegraphics[width=0.48\textwidth]{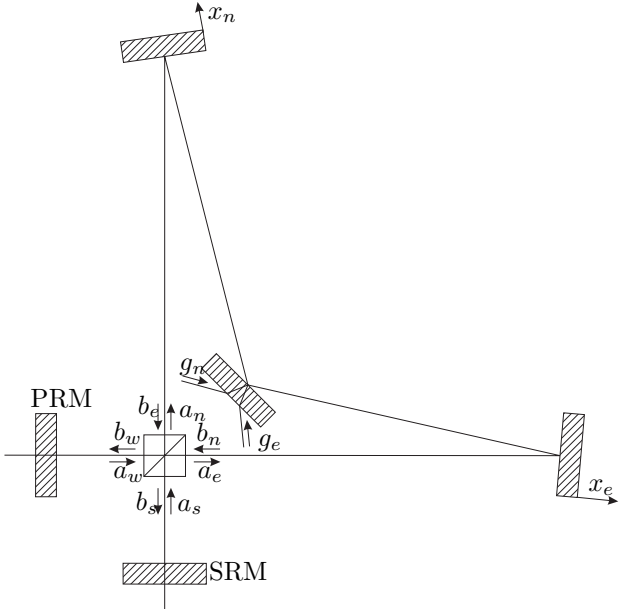}\\
\caption{Power-
and signal-recycled speed meter interferometer.}\label{fig3a}
\end{center}
\end{wrapfigure}

Then we need to express $A$ in terms of $A'$, the amplitude of the beam $a'$ entering the PRM. One can write down the following equations:
\begin{subequations}
\begin{equation}
A''=it_{PR}A'-r_{PR}B''\,,\quad A''=Ae^{-i\phi_{PR}}\,,
\end{equation}
\begin{equation}
B'=it_{PR}B''-r_{PR}A''\,,\quad B''=Be^{i\phi_{PR}}\,,
\end{equation}
\begin{equation}
B=i(1-\alpha_{loss})A\,,
\end{equation}
\end{subequations}
where $t_{PR}$ and $r_{PR}$ are PRM transmittance and reflectivity, $\phi_{PR}=\omega_\circ L/c$, and $\alpha_{loss}$ represents losses in the entire scheme (we will neglect these losses in this calculation as  $\alpha_{loss}$ is sufficiently small). Solving these equations leads to the following expression:
\begin{equation}\label{PRMsol}
A=\dfrac{it_{PR}e^{i\phi_{PR}}}{1+ir_{PR}e^{2i\phi_{PR}}}A'\,.
\end{equation}
In order to maximize circulating power one should tune PR-cavity so that $\phi_{PR}=\pi/4$, then we will obtain that
\begin{equation}
W_{PR}=\dfrac{t_{PR}^2 W_\circ}{(1-r_{PR})^2}\simeq\dfrac{4W_\circ}{t^2_{_PR}}\,,
\end{equation}
where $W_\circ=\hbar\omega_\circ|A'|$, and $W_{PR}=\hbar\omega_\circ|A|^2$.

\paragraph{Signal recycling mirror.}
In the case of signal recycling cavity only quantum fluctuations
transformation is worth to be examined. As the interferometer is
tuned so that output port is kept "dark", classical amplitude of
the beam leaving the BS is equal to zero. Now $a$ and $b$ stand
for fluctuations entering and leaving the BS, while $a'$ and $b'$
stand for fluctuations entering and leaving the SRM. Therefore we
write down the following equations:
\begin{equation}
\hat b'=it_{SR}e^{i\phi_{SR}}\hat b-r_{SR}\hat a'\,,\quad \hat a'e^{-i\phi_{SR}}=it_{SR}\hat a'-r_{SR}e^{i\phi_{SR}}\hat b\,,
\end{equation}
where $t_{SR}$ and $r_{SR}$ are SRM transmittance and
reflectivity, $\phi_{SR}=\omega_\circ L/c$. In these equations we
should suppose $\hat a'\equiv \hat a_s$, $\hat b'\equiv \hat b_s$,
and the equation that gives the relation between these operators
is represented by expression (\ref{SimpleOutput}).

Solution of these equations gives us the following expression for
PR\&SR speed meter output signal:
\begin{equation}\label{SRSMIOrels}
\hat b_s'=-\left(r_{SR}+\dfrac{t_{SR}^2e^{2i\phi_{SR}}\beta_{input}}{1+r_{SR}\beta_{input}e^{2i\phi_{SR}}} \right)\hat a_s'+\dfrac{it_{SR}e^{i\phi_{SR}}\beta_{loss}}{1+r_{SR}\beta_{input}e^{2i\phi_{SR}}}\hat g_s+{\cal K}_{SR}x_-\,,
\end{equation}
where
$${\cal K}_{SR}=\dfrac{it_{SR}e^{i\phi_{SR}}\beta_{loss}}{1+r_{SR}\beta_{input}e^{2i\phi_{SR}}}{\cal K}_{simple}\,,$$
$\beta_{input}$, $\beta_{loss}$, and ${\cal K}_{simple}$ are defined in (\ref{SimpleOutput}). The value of $\phi_{SR}$  is chosen so that output signal is maximal, \textit{i. e.} ${\cal K}_{SR}\rightarrow\ max$. Therefore, signal recycling cavity should be tuned in the way to provide $\phi_{SR}=\pi/4$.

Now we are able to calculate spectral densities of radiation pressure and shot noises and their cross-correlation spectral density.

\subsection{Radiation pressure noise spectral density $S_F(\Omega)$}
In order to calculate radiation pressure noise spectral density
for power- and signal-recycled speed meter interferometer one
needs to replace operator $\hat a_s$ in formula (\ref{Ffluct}) of
previous appendix by operator $\hat a'_s$ that can be expressed in
terms of $\hat a_s$ and $\hat g_s$ as
\begin{equation}
  \hat
  a'_s=\dfrac{it_{SR}e^{i\phi_{SR}}\hat
  a_s-r_{SR}\beta_{loss}e^{2i\phi_{SR}}\hat
  g_s-r_{SR}e^{2i\phi_{SR}}{\cal
  K}_{simple}x_-}{1+r_{SR}\beta_{input}e^{2i\phi_{SR}}}\,.
\end{equation}
After the same operations as in Appendix \ref{SimpleSMSf} one will obtain that radiation pressure noise spectral density in case of power- and signal-recycled speed meter interferometer is equal to
\begin{equation}\label{SRSMSf}
S_F(\Omega)=\dfrac{16\hbar\omega_\circ\tau W_{PR}}{c^2}\cdot\dfrac{(\gamma+\gamma_{SR})(\gamma\gamma_{SR}+\Omega^2)}{(\gamma+\gamma_{SR})^2+4\Omega^2}\,.
\end{equation}

\subsection{Shot noise spectral density $S_x(\Omega)$}
To obtain the expression for shot noise spectral density in the case of signal-recycled interferometer we should use the same procedure as in Appendix \ref{SimpleSMSx} but for one exception. We should substitute value of ${\cal K}_{simple}$ by ${\cal K}_{SR}$. To account for power recycling we should also write down $W_{PR}$ instead of $W$. After all we shall obtain the following formula:
\begin{equation}\label{SRSMSx}
S_x(\Omega)=\dfrac{\hbar c^2}{64 \omega_\circ\tau W_{PR}\sin^2\Psi}\cdot\dfrac{(\gamma+\gamma_{SR})^2+4\Omega^2}{\gamma_{SR}(\gamma^2+\Omega^2)}\,.
\end{equation}

\subsection{Cross-correlation spectral density $S_{xF}(\Omega)$}
Cross-correlation spectral density in this case is the same as in previous section and is common for all considered interferometric schemes with homodyne detection:
\begin{equation}\label{SRSMSxf}
  S_{xF}(\Omega)=-\frac\hbar2\cot\Psi\,.
\end{equation}

\subsection{Power- and signal-recycled speed meter sensitivity}
Total noise of the PR\&SR speed meter is described by the same formula as (\ref{simplStotal}). Spectral densities of noises in this particular case can be obtained from formulae (\ref{SRSMSf}), (\ref{SRSMSx}), and (\ref{SRSMSxf}).

Being substituted to (\ref{simplStotal}) and divided by $S_{SQL}=\hbar m\Omega^2$ these formulae will give us the expression for the factor $\xi^2$ by which one can beat the SQL using PR\&SR speed meter:
\begin{equation}\label{SRSMxi}
\xi^2=P_{SR}\dfrac{(\gamma_{SR}+\gamma)(\gamma\gamma_{SR}+\Omega^2)}{\Omega^2((\gamma+\gamma_{SR})^2+4\Omega^2)}+\dfrac{1+\cot^2\Psi}{4P_{SR}}\cdot\dfrac{\Omega^2((\gamma+\gamma_{SR})^2+4\Omega^2)}{\gamma_{SR}(\gamma^2+\Omega^2)}+\cot\Psi\,,
\end{equation}
where $P_{SR}=\dfrac{16\omega_\circ\tau W_{PR}}{mc^2}$, $\gamma=\dfrac{1-r}{2\tau}$ is the interferometer half-bandwidth part due to optical losses and $\gamma_{SR}=\dfrac{1-r_{SR}}{2\tau}$ is the part of half-bandwidth due to signal-recycling mirror. We can now optimize this expression at some fixed frequency $\Omega_\circ$ with respect to homodyne angle $\Psi$, and circulating power $W_{PR}$.
The optimal circulating power $W_{opt}$ and homodyne angle $\Psi_{opt}$ for considered scheme are equal to:
\begin{equation}
W_{opt}\simeq\dfrac{mc^2}{32\omega_\circ\tau}\left(1+4\dfrac{\Omega^2_\circ}{\gamma_{SR}^2} \right)\sqrt{\dfrac{\Omega_\circ^2}{\gamma\gamma_{SR}}}\,,\quad\Psi_{opt}\simeq-\arctan{\sqrt{\dfrac{\gamma\gamma_{SR}}{\Omega_\circ^2}}} \,,
\end{equation}
and the minimal value of $\xi$ that can be achieved at frequency $\Omega_\circ$ is
\begin{equation}\label{SRSMxiopt}
\xi_{opt}\simeq\sqrt[4]{\dfrac{\gamma\gamma_{SR}}{\Omega_\circ^2}}\,.
\end{equation}
We can see that to beat the SQL considerably using signal-recycled speed meter one needs to decrease internal losses. Decreasing $\gamma_{SR}$ (or increasing $r_{SR}$) will also increase the scheme sensitivity at given frequency but  at the sacrifice of sensitivity at other frequencies.

Let us estimate the optimal circulating power that is necessary to obtain $\xi\simeq0.1$ . Let substitute the following parameters:
\begin{equation}
\Omega_\circ=10^3 \mbox{ s}^{-1}\,,\,m=5 \mbox{ kg}\,,\,L=600\mbox{ m}\,,\,\omega_\circ=1.77\cdot10^{15}\mbox{ s}^{-1}\,,\,1-r=10^{-5}\,,
\end{equation}
that are typical for GEO 600 interferometer. To obtain $\xi_{opt}=0.1$ one needs $\gamma_{SR}=40\mbox{ s}^{-1}$ ($1-r_{SR}=1.6\cdot10^{-4}$) and circulating power of the order of
\begin{equation}
W_{opt}=10^{11}\mbox{ W}\,.
\end{equation}
We can see that considered scheme of power- and signal-recycled speed meter can be hardly implemented in gravitational wave detection, as it requires enormous amount of optical power and its sensitivity is high only in very narrow frequency band.

This result confirms the statement we have made in section \ref{SRPMvsSRSM}, \textit{i. e.} in order to achieve the same sensitivity as position meter with Fabry-Perot cavities in arms, speed meter requires highly reflecting SRM to be used and therefore enormous amount of pumping power is needed. Taking it into account it is reasonable to conclude that the scheme with Fabry-Perot cavities in both arms without SRM looks like the best candidate for implementing as speed meter interferometer.

\section{Derivation of spectral densities for speed meter
interferometer with Fabry-Perot cavities in arms}\label{App2}
\subsection{Fabry-Perot input-output relations}
To obtain input-output (IO) relations for the scheme presented in
Fig. \ref{fig4} one needs to know IO relations for Fabry-Perot
cavity with movable end mirror. The case of two movable mirrors
will give the same result but when investigating cavity dynamics
one should replace end mirror displacement $x$ by the difference
$x_1-x_2$ of each mirror displacements, and mirrors masses should
be taken one half of real value to account for changes in
radiation pressure forces.

\begin{wrapfigure}[8]{l}{65mm}

\psfrag{A1}[lb]{$a_1$}\psfrag{B1}[lb]{$b_1$}
\psfrag{C1}[lb]{$c_1$}\psfrag{D1}[lb]{$d_1$}\psfrag{C2}[lb]{$c_2$}\psfrag{D2}[lb]{$d_2$}
\psfrag{R1}[cb]{$-r_1,\,it_1,\,ia_1$}\psfrag{R2}[cb]{$-r_2,\,0,\,ia_2$}\psfrag{X}[lb]{$x$}
\psfrag{L}[lb]{$L=c\tau$}
\includegraphics[width=60mm]{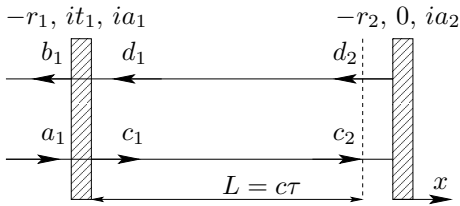}
\caption{Fabry-Perot cavity with movable mirror}\label{fig6}

\end {wrapfigure}

Consider FP cavity presented in Fig. \ref{fig6}. We suppose end
mirror to have zero transmittance for the same reason as in
subsection \ref{SubSec31}.

Suppose we know classical amplitude of internal field $C_1$ and
input quantum fluctuations $\hat a_1$ of input light. To obtain
expressions for internal and output fields one needs to solve the
following equations:

\begin{equation}
\begin{array}{ccc}
  &\hat b_1(\omega)=-r_1\hat a_1(\omega)+it_1\hat c_1(\omega)+ia_1\hat
  g_{12}(\omega)\,,\\
  \\
  &\hat d_1(\omega)=it_1\hat a_1(\omega)-r_1\hat c_1(\omega)+ia_1\hat
  g_{11}(\omega)\,,\\
  \\
  &\hat d_2(\omega)=-r_2\hat c_2(\omega)+ia_2\hat
  g_{21}(\omega)-2i\kappa(\omega)r_2C_2x(\omega_\circ
  -\omega)\,,\\
  \\
  &\hat c_2(\omega)=\hat d_1(\omega)e^{i\omega\tau}\,,\quad \hat d_2(\omega)=\hat
  c_1(\omega)e^{-i\omega\tau}\,.
\end{array}
\end{equation}
The solution can be written as:
\begin{subequations}\label{FPIO}
\renewcommand{\theequation}{\theparentequation.\arabic{equation}}
\begin{multline}\label{b1}
  \hat b_1=\frac{1}{{\cal
L}(\omega)}[{\cal B}(\omega)\hat
a_1(\omega)-e^{2i\omega\tau}r_2t_1a_1\hat
g_{11}(\omega)+\\+ia_1{\cal L}(\omega)\hat
g_{12}(\omega)+e^{i\omega\tau}t_1a_2\hat
g_{21}(\omega)-2\kappa(\omega)e^{i\omega\tau}r_2t_1C_2x(\omega_\circ
-\omega)]\,,
\end{multline}
\begin{equation}
   \hat c_2=\frac{1}{{\cal
L}(\omega)}[-ie^{i\omega\tau}t_1\hat
a_1(\omega)-ie^{i\omega\tau}a_1\hat
g_{11}(\omega)+ie^{2i\omega\tau}r_1a_2\hat
g_{21}(\omega)-2i\kappa(\omega)e^{2i\omega\tau}r_1r_2C_2x(\omega_\circ
-\omega)]\,,
\end{equation}
\begin{equation}
   \hat d_2=\frac{1}{{\cal
L}(\omega)}[ie^{i\omega\tau}t_1r_2\hat
a_1(\omega)+ie^{i\omega\tau}r_2a_1\hat g_{11}(\omega)-ia_2\hat
g_{21}(\omega)+2i\kappa(\omega)r_2C_2x(\omega_\circ -\omega)]\,,
\end{equation}
\end{subequations}
where
\begin{equation}
  {\cal L}(\omega)=r_1r_2e^{2i\omega\tau}-1\,,\quad {\cal
  B}(\omega)=r_1-e^{2i\omega\tau}r_2(r_1^2+t_1^2)\,.
\end{equation}
Expressions (\ref{FPIO}) represent the FP-cavity input-output
relations.
\subsection{Radiation pressure noise in FP-cavity}\label{subsecFPradpress}
In this subsection we will obtain the expression for radiation
pressure force acting upon the movable mirror of the FP-cavity. In
accordance with formula for radiation pressure we can write down
the following:
\begin{equation}
 \hat F_{r.p.}=\frac{\hat W_{c_2}+\hat W_{d_2}}{c}=\frac{\overline{\hat E_{c_2}^2}+\overline{\hat E_{d_2}^2}}{4\pi}{\cal A}\,.
\end{equation}
Taking into account expressions (\ref{FPIO}) one can obtain:
\begin{multline}\label{FPradpress}
  \hat
  F(t)=F_\circ +\hbar\int_0^\infty\kappa(\omega)[C_2^*\hat c_2(\omega)+D_2^*\hat d_2(\omega)]e^{i(\omega_\circ -\omega)t}\frac{d\omega}{2\pi}+{\rm
  h.\,c.}=\\=F_\circ +\hbar\int_0^\infty\kappa(\omega)[F_{a_1}\hat a_1(\omega)+F_{g_{11}}\hat g_{11}(\omega)+F_{g_{21}}\hat g_{21}(\omega)]e^{i(\omega_\circ -\omega)t}\frac{d\omega}{2\pi}+{\rm
  h.\,c.}\,,
\end{multline}
where
\begin{equation}
 F_{a_1}\simeq-\frac{2it_1e^{i\omega\tau}C_2^*}{{\cal
L}(\omega)}\,,\quad
F_{g_{11}}\simeq-\frac{2ia_1e^{i\omega\tau}C_2^*}{{\cal
L}(\omega)}\,,\quad
F_{g_{21}}\simeq\frac{ia_2(e^{2i\omega\tau}+1)C_2^*}{{\cal
L}(\omega)}\,,
\end{equation}
$F_\circ=\dfrac{\hbar\omega_\circ|C_2|^2(1+r_2^2)}{c}\simeq\dfrac{2\hbar
\omega_\circ|C_2|^2}{c}$ is the constant classical radiation
pressure force. Items that correspond to other noise operators are
neglected as they are small compared to the above values.

\subsection{Input-output relations derivation}
Now we can write down IO relations for our speed meter scheme. One
can notice that light beam between input and output moments is
sequentially reflected from two FP cavities. Therefore, sideband
operator that describes output beam for the first cavity at the
same time describes input beam for the second one. Moreover, the
beam that enters cavity for the first time and the beam that
enters cavity being once reflected do not interact as they have
different polarizations ($\circlearrowleft$ and
$\circlearrowright$ in our case). Then, beams that leaves the
scheme falling to the beam splitter from the "north" and "east"
are characterized by sideband operators $\hat c_n$ and $\hat c_e$
that can be expressed in terms of input beams operators as
\begin{subequations}\label{SMIO}
\renewcommand{\theequation}{\theparentequation.\arabic{equation}}
\begin{multline}
  \hat c_{n}=\frac{1}{{\cal
L}(\omega)}[{\cal B}(\omega)\hat
b_{n}^I(\omega)-e^{2i\omega\tau}r_2t_1a_1\hat
g_{e_{11}}^{II}(\omega)+e^{i\omega\tau}t_1a_2\hat
g_{e_{21}}^{II}(\omega)+i\sqrt2\kappa(\omega)e^{i\omega\tau}r_2t_1Fx_e(\omega_\circ
-\omega)]\,,
\end{multline}
\begin{multline}
  \hat c_{e}=\frac{1}{{\cal
L}(\omega)}[{\cal B}(\omega)\hat
b_{e}^I(\omega)-e^{2i\omega\tau}r_2t_1a_1\hat
g_{n_{11}}^{II}(\omega)+e^{i\omega\tau}t_1a_2\hat
g_{n_{21}}^{II}(\omega)-\sqrt2\kappa(\omega)e^{i\omega\tau}r_2t_1Fx_n(\omega_\circ
-\omega)]\,,
\end{multline}
\end{subequations}
where $F$ is the amplitude inside the cavity during second
reflection, $\hat b_{1_n}^{I}$ and $\hat b_{1_e}^{I}$ are defined
by formula (\ref{b1}) with replacing index $_1$ by $_{1_n}^{I}$
and $_{1_e}^{I}$ (superscript $^{I}$ means first reflection and
superscript $^{II}$ means second reflection from FP cavity), and
$C_2$ by $E_n=iE/\sqrt2$ and $E_e=-E/\sqrt2$, where where $E$ is
the amplitude inside the cavity during the first reflection. Here
we also introduced additional pumping to compensate energy losses
due to absorption in FP cavities. We suppose that this pumping
should feed the main beam before it enters the second cavity, then
parameter $\pmb{\eta}=F_e/E_e=F_n/E_n$. Now we are able to
calculate output beam sideband operator $\hat c_s=\dfrac{i\hat
c_n-\hat c_e}{\sqrt2}$
\begin{multline}\label{SMoutApp}
  \hat c_{s}= \frac{1}{{\cal L}^2(\omega)}[i{\cal
B}^2(\omega)\hat a_{s}^I(\omega)+i{\cal
B}(\omega)(e^{2i\omega\tau}r_2t_1a_1\hat
g_{s_{11}}^I(\omega)-e^{i\omega\tau}t_1a_2\hat
g_{s_{21}}^{I}(\omega))-\\-i{\cal
L}(\omega)(e^{2i\omega\tau}r_2t_1a_1\hat
g_{w_{11}}^{II}(\omega)-e^{i\omega\tau}t_1a_2\hat
g_{w_{21}}^{II}(\omega))-({\cal B}_1(\omega)-\pmb{\eta}{\cal
L}(\omega))E\kappa(\omega)e^{i\omega\tau}r_2t_1x_-(\omega_\circ
-\omega)]\,.
\end{multline}
where $x_-=\dfrac{x_n-x_e}{2}$, $\hat \alpha_s=-\dfrac{\hat
\alpha_n+i\hat \alpha_e}{\sqrt2}$, $\hat \alpha_w=-\dfrac{\hat
\alpha_e+i\hat \alpha_n}{\sqrt2}$ (here $\hat \alpha$ stands for
any bosonic operator).

\subsection{Radiation pressure noise spectral density $S_F(\Omega)$}
In order to calculate the radiation pressure noise for the speed
meter as a whole we need to use the expression for radiation
pressure fluctuational force acting upon the end mirror of
Fabry-Perot cavity, presented in subsection \ref{subsecFPradpress}
by formula (\ref{FPradpress}). As in simple scheme one can present
fluctuational force acting upon each of FP-cavity end mirrors as a
sum of two independent items:
\begin{equation}\label{SMradpressforce_e/n}
  \hat F_{e}=\hat F_e^{\circlearrowright}+\hat F_e^{\circlearrowleft},\quad\mbox{and}\quad
  \hat F_{n}=\hat F_n^{\circlearrowright}+\hat F_n^{\circlearrowleft}\,,
\end{equation}
and the net force acting upon the scheme can be presented as
\begin{equation}\label{SMradpressforce}
  \hat F=\hat F^{\circlearrowright}+\hat F^{\circlearrowleft}\,,
\end{equation}
where
\begin{equation}
  \hat F^{\circlearrowright}=\hat F_n^{\circlearrowright}-\hat F_e^{\circlearrowright},\quad\mbox{and}\quad
  \hat F^{\circlearrowleft}=\hat F_n^{\circlearrowleft}-\hat
  F_e^{\circlearrowleft}\,.
\end{equation}
Let write down the explicit form of these expressions
\begin{subequations}
\renewcommand{\theequation}{\theparentequation.\arabic{equation}}
\begin{multline}
  \hat F^{\circlearrowright}(t)=F_\circ ^{\circlearrowright}+2\hbar\int_0^\infty\kappa(\omega)[F^{\circlearrowright}_{a_{s}}\hat a_{s}(\omega)+F^{\circlearrowright,\,I}_{g_{s_{11}}}\hat g_{s_{11}}^I(\omega)+F^{\circlearrowright,\,I}_{g_{s_{21}}}\hat g_{s_{21}}^I(\omega)]e^{i(\omega_\circ -\omega)t}\frac{d\omega}{2\pi}+{\rm
  h.\,c.}\,,
\end{multline}
where $F_\circ
^{\circlearrowright}\simeq\dfrac{\hbar\omega_\circ|E|^2}{c}$ is
the corresponding classical radiation pressure force,
\begin{equation}
F^{\circlearrowright}_{a_{s}}\simeq\frac{2t_1e^{i\omega\tau}
E^*}{{\cal L}(\omega)}\,,
\end{equation}
\begin{equation}
F^{\circlearrowright,\,I}_{g_{s_{11}}}\simeq\frac{2a_1e^{i\omega\tau}
E^*}{{\cal L}(\omega)},\quad
F^{\circlearrowright,\,I}_{g_{s_{21}}}\simeq-\frac{ia_2(e^{2i\omega\tau}+1)
E^*}{{\cal L}(\omega)}\,,
\end{equation}
\end{subequations}
and
\begin{subequations}
\renewcommand{\theequation}{\theparentequation.\arabic{equation}}
\begin{multline}
  \hat F^{\circlearrowleft}(t)=F_\circ ^{\circlearrowleft}+2\hbar\int_0^\infty\kappa(\omega)[F^{\circlearrowleft}_{a_{s}}\hat a_{s}(\omega)+F^{\circlearrowleft,\,I}_{g_{s_{11}}}\hat g_{s_{11}}^I(\omega)+F^{\circlearrowleft,\,I}_{g_{s_{21}}}\hat g_{s_{21}}^I(\omega)+\\+F^{\circlearrowleft,\,II}_{g_{s_{11}}}\hat g_{s_{11}}^{II}(\omega)+F^{\circlearrowleft,\,II}_{g_{s_{21}}}\hat g_{s_{21}}^{II}(\omega)]e^{i(\omega_\circ -\omega)t}\frac{d\omega}{2\pi}+{\rm
  h.\,c.}\,,
\end{multline}
where $F_\circ
^{\circlearrowleft}\simeq\dfrac{\hbar\omega_\circ\eta^2|E|^2}{c}$
is the corresponding classical radiation pressure force,
\begin{equation}
F^{\circlearrowleft}_{a_{s}}\simeq-\frac{2t_1e^{i\omega\tau}
\pmb{\eta}^*E^*}{{\cal L}(\omega)}\cdot\frac{{\cal
B}(\omega)}{{\cal L}(\omega)}\,,
\end{equation}
\begin{equation}
F^{\circlearrowleft,\,I}_{g_{s_{11}}}\simeq\frac{2a_1e^{i\omega\tau}
\pmb{\eta}^*E^*}{{\cal
L}(\omega)}\cdot\frac{t_1a_1e^{2i\omega\tau}}{{\cal
L}(\omega)},\quad
F^{\circlearrowleft,\,I}_{g_{s_{21}}}\simeq\frac{ia_2(e^{2i\omega\tau}+1)
\pmb{\eta}^*E^*}{{\cal
L}(\omega)}\cdot\frac{t_1a_2e^{i\omega\tau}}{{\cal L}(\omega)}\,,
\end{equation}
\begin{equation}
F^{\circlearrowleft,\,II}_{g_{s_{11}}}\simeq-\frac{2a_1e^{i\omega\tau}
\pmb{\eta}^*E^*}{{\cal L}(\omega)},\quad
F^{\circlearrowleft,\,II}_{g_{s_{21}}}\simeq-\frac{ia_2(e^{2i\omega\tau}+1)
\pmb{\eta}^*E^*}{{\cal L}(\omega)}\,,
\end{equation}
\end{subequations}
Spectral density of radiation pressure noise can be calculated
using the same technique as in Appendix \ref{App1}, {\it i. e.}
\begin{equation}\label{SMSfgeneral}
  S_F(\Omega)=\frac12(S_F'(\omega_\circ-\Omega)+S_F'(\omega_\circ+\Omega))\,,
\end{equation}
where
\begin{multline}
S_F'(\omega_\circ-\Omega)=4\hbar^2\kappa^2(\omega_\circ-\Omega)(|F^{\circlearrowright}_{a_{s}}+F^{\circlearrowleft}_{a_{s}}|^2+|F^{\circlearrowright,\,I}_{g_{s_{11}}}+F^{\circlearrowleft,\,I}_{g_{s_{11}}}|^2+|F^{\circlearrowright,\,I}_{g_{s_{21}}}+F^{\circlearrowleft,\,I}_{g_{s_{21}}}|^2+\\+|F^{\circlearrowleft,\,II}_{g_{s_{11}}}|^2+|F^{\circlearrowleft,\,II}_{g_{s_{21}}}|^2)\,,
\end{multline}
If one uses formulae (\ref{NBconditions1}) and
(\ref{NBnotations2}), then it is easy to obtain that radiation
pressure noise spectral density in narrow-band approximation is
defined by the following expression:
\begin{equation}\label{SMSf}
  S_{F}(\Omega)=\frac{8\hbar\omega_\circ\tau
  W}{L^2}\frac{\gamma(1+\eta^2)(\gamma^2+\Omega^2)-2\gamma_1\eta(\gamma^2-\Omega^2)}{(1+\eta^2)(\gamma^2+\Omega^2)^2}\,,
\end{equation}
where $W=\hbar\omega_\circ \dfrac{|E|^2(1+\eta^2)}{8}$ is the
light power at the end mirror.

\subsection{Shot noise spectral density $S_x(\Omega)$}
Shot noise spectral density can be obtained in the same manner as
in Appendix \ref{App1}. Here
\begin{equation}
  K(\Omega) = -\frac{2i\omega_\circ ^{3/2}|E|}{c}\frac{e^{-i\Omega\tau}r_2t_1}{{\cal L}^2(\omega_\circ -\Omega)}[{\cal
  B}_1(\omega_\circ -\Omega)\sin\Psi-\eta
  {\cal L}(\omega_\circ -\Omega)\sin(\Phi+\Psi)]\,,
\end{equation}
where $\Psi=\phi_{\rm LO}+\arg C$, $\eta=|\pmb{\eta}|$,
$\Phi=\arg\pmb{\eta}$, and spectral density is defined by the
following expression:
\begin{equation}\label{SMSxApp}
  S_x(\Omega)=\frac{\omega_\circ}{|K(\Omega)|^2}=\frac{\hbar c^2}{32\omega_\circ W}\frac{(1+\eta^2)|{\cal L}(\omega_\circ -\Omega)|^2}{r_2^2t_1^2|{\cal
  B}_1(\omega_\circ -\Omega)\sin\Psi-\eta
  {\cal L}(\omega_\circ -\Omega)\sin(\Phi+\Psi)|^2}\,.
\end{equation}

Using the narrow band approximation defined by
(\ref{NBconditions1}) and (\ref{NBnotations2}) the following
expression for $S_x$ can be written if one suppose $\Phi=0$:
\begin{equation}\label{SMSxnarrow}
  S_x(\Omega)=\frac{\hbar
  L^2}{32\omega_\circ \tau W}\frac{(1+\eta^2)(\gamma^2+\Omega^2)^2}{\gamma_1\sin^2\Psi(((1+\eta)\gamma-2\gamma_1)^2+(1+\eta)^2\Omega^2)}\,,
\end{equation}
where $L=c\tau$.

\subsection{Real scheme cross-correlation spectral
density $S_{xF}$} It can be shown that cross-correlation spectral
density for the real speed meter scheme with optical losses is the
same as for the ideal one, \textit{i. e.}
\begin{equation}\label{SMSxf2}
  S_{xF}=-\dfrac{\hbar}{2}\cot\Psi\,.
\end{equation}
The above expression does not depend on frequency and is the same
in narrow band approximation.

\section{Narrow-band optimization of $\xi^2$}\label{App3}
In this section we will perform the optimization of speed meter interferometer sensitivity at some fixed given frequency $\Omega_\circ$. In this connection it seems convenient to introduce new dimensionless variables
\begin{equation}
\upsilon=\dfrac{\Omega}{\gamma}\,,\quad\varepsilon=\dfrac{\gamma-\gamma_1}{\gamma}=\dfrac{\alpha}{\gamma}\,,\quad P=\dfrac{16\omega_\circ\tau W}{mL^2\gamma^3}\,,\quad A=\cot\Psi\,,
\end{equation}  
to rewrite (\ref{SMxi}) as 
\begin{equation}\label{SMxi_dimensionless}
\xi^2=\frac12\left[Pa+\frac{1+A^2}{P}b\right]+A\,, 
\end{equation} 
where 
\begin{equation}
a=\frac{\varepsilon+\upsilon^2(2-\varepsilon)}{\upsilon^2(1+\upsilon^2)^2}\,,\quad b=\frac{\upsilon^2(1+\upsilon^2)^2}{2(1-\varepsilon)(\varepsilon^2+\upsilon^2)}\,.
\end{equation} 
Optimizing (\ref{SMxi_dimensionless}) with respect to $P$ and $A$ one can readily show that it reaches minimum at 
\begin{equation}
P=\frac{b}{\sqrt{ab-1}}\,,\mbox{ and } A=-\frac{1}{\sqrt{ab-1}}\,,
\end{equation} 
where $a$ and $b$ should be taken at frequency $\Omega_\circ$.
Being substituted to  (\ref{SMxi_dimensionless}) these expressions will turn it to
\begin{equation}\label{xi_opt_intermed}
\xi^2=\sqrt{ab-1}\,.
\end{equation} 
The second step is the minimization of the above expression with respect to $\gamma$. Obviously, to obtain the optimal value of $\gamma$ it is necessary to solve the following equation
$$\dfrac{\partial K}{\partial\gamma}=0\,,$$
where $K=ab$. Here we should remember that $\varepsilon=\alpha/\gamma$ and $\upsilon=\Omega_\circ/\gamma$, then the above equation will be trasformed to
$$ \varepsilon\dfrac{\partial K}{\partial\varepsilon}+\upsilon\dfrac{\partial K}{\partial\upsilon}=0\,, $$
that after simplification will be written as
\begin{equation}
\upsilon^2+2\varepsilon-1=\dfrac{\Omega_\circ^2}{\gamma^2}+2\dfrac{\alpha}{\gamma}-1=0\,.
\end{equation}  
The positive solution of this equation is
\begin{equation}
\gamma_{opt}=\alpha+\sqrt{\alpha^2+\Omega_\circ^2}\simeq\Omega_\circ\,,
\end{equation} 
where the last approximate equality corresponds to the case of small losses $\alpha\ll\Omega_\circ$. 
Substituting the obtained results to (\ref{xi_opt_intermed}) with respect to the case of small losses, one will have
\begin{equation}
\xi_{min}=\sqrt[4]{\dfrac{\varepsilon+\varepsilon^2-\varepsilon^3}{1-\varepsilon+\varepsilon^2-\varepsilon^3}}\simeq\sqrt[4]{\varepsilon}=\sqrt[4]{\dfrac{\alpha}{\Omega_\circ}}\,,
\end{equation}  
that is the same expression that is presented in formula (\ref{xi_min}).

\end{document}